\documentclass{siamltex}

\usepackage{amssymb}

\usepackage{amsmath}

\usepackage[left=2cm,right=2cm,top=2cm,bottom=2cm]{geometry}

\usepackage{color}

\usepackage{graphicx}

\usepackage{enumitem}

\usepackage{float}

\newcommand{\bq}{{\bf q}}
\newcommand{\bp}{{\bf p}}
\newcommand{\bb}{{\bf b}}
\newcommand{\eps}{\varepsilon}

\title{A Discrete-to-Continuum Model of Weakly Interacting Incommensurate Two-Dimensional Lattices}

\author{
    Malena I.~Espa\~ nol\footnotemark[1]
\and
    Dmitry Golovaty\footnotemark[1]
\and
    J. Patrick Wilber\footnotemark[1]}

\begin{document}
\maketitle

\renewcommand{\thefootnote}{\fnsymbol{footnote}}
\footnotetext[1]{
    Department of Mathematics,
    University of Akron,
    Akron,
    OH 44325, USA.}

\begin{abstract}
In this paper we propose a continuum variational model for a
two-dimensional deformable lattice of atoms interacting with a
two-dimensional rigid lattice.  The two lattices have slightly
different lattice parameters and there is a small relative rotation
between them.  This is a prototypical example of a three-dimensional
system consisting of a graphene sheet suspended over a substrate. The
continuum model recovers both qualitatively and quantitatively the
behavior observed in the corresponding discrete model.  The continuum
model predicts that the deformable lattice develops a network of
domain walls characterized by large shearing,
stretching, and bending deformation that accommodate the misalignment
and/or mismatch between the deformable and rigid lattices.  Two integer-valued parameters that can be identified with the components of a Burgers vector, describe the mismatch
between the lattices and determine the geometry and the detail of
deformation associated with the domain walls.
\end{abstract}

\section{Introduction} \label{s1}

The mechanical properties of interacting layers of two-dimensional
crystals are currently a topic of intense investigation.  Bilayer
graphene is perhaps the most notable motivation for these studies.
Other motivating examples include few layer crystals of hexagonal
boron nitride (h-BN), molybdenum disulphide (MoS$_2$), and tungsten
diselenide (WSe$_2$).  More generally, there is an interest in
modeling and simulating the mechanical properties of van der Waals
heterostructures, a term that describes stacks of possibly different
two-dimensional crystals \cite{geim2013van,novoselov20162d}.  Interest
in these heterostructures is currently driven by the idea that it may be
possible to engineer advanced materials with novel properties by
stacking different types of individual layers in appropriate
sequences.

For bilayer graphene and other interacting layers of two-dimensional
crystals, the deformation of the bilayer is determined by the strong,
bonded interactions between nearest neighbors on a given layer and by
the weak, non-bonded interactions between nearby atoms on different
layers.  The weak interactions, although sufficient to hold the layers
together, permit sliding and rotations between the layers.  Hence if
the layers have different lattice geometries or lattice parameters or
if a bilayer is synthesized with local regions in different stacking
arrangements, the layers may adjust by shifting or rotating locally.

These adjustments can induce strain within each layer that can be
relaxed by both in-plane and out-of-plane, atomic-scale displacements
of the atoms on the layers.  What typically occurs for layers with
slightly mismatched lattices is that the atomic-scale displacements
create relatively large commensurate regions separated by localized
incommensurate regions.  In the commensurate regions the interlayer
energy is minimized.  In the localized incommensurate regions, or
domain walls, strain may be relaxed by out-of-plane displacement.
These displacements can generate interesting larger-scale pattern
formation that may strongly influence the electrical, thermal, and
other properties of the bilayer \cite{geim2013van,PhysRevB.91.205412}.

\begin{figure}[htb]
\centering
    \includegraphics[width=.5\linewidth]
                    {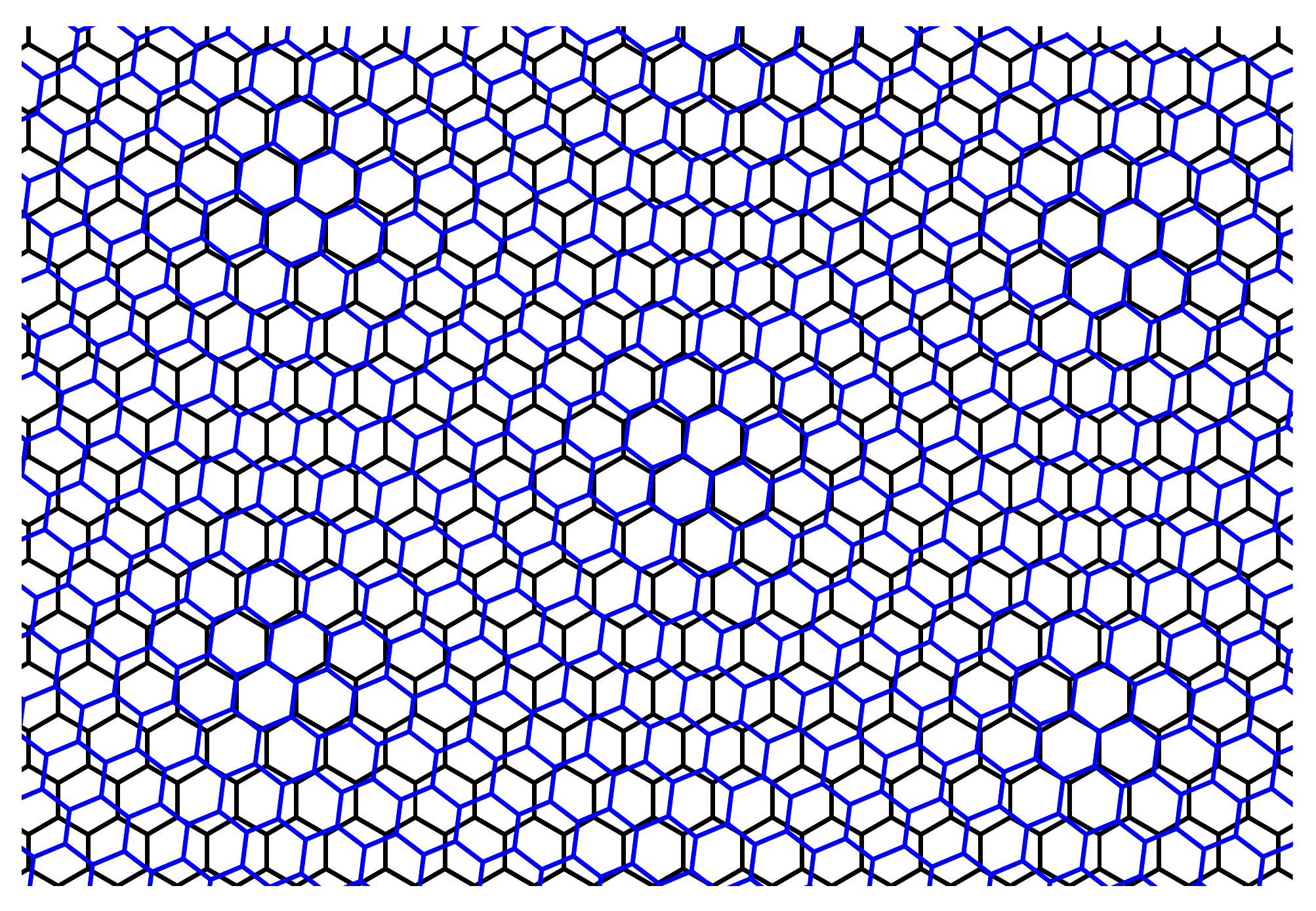}
  \caption{Moir\'e pattern in slightly misoriented hexagonal lattices with relative rotation $7.2^\circ$.}
  \label{moire_fig_hex}
\end{figure}

\begin{figure}[htb]
\centering
    \includegraphics[width=.3\linewidth]
                    {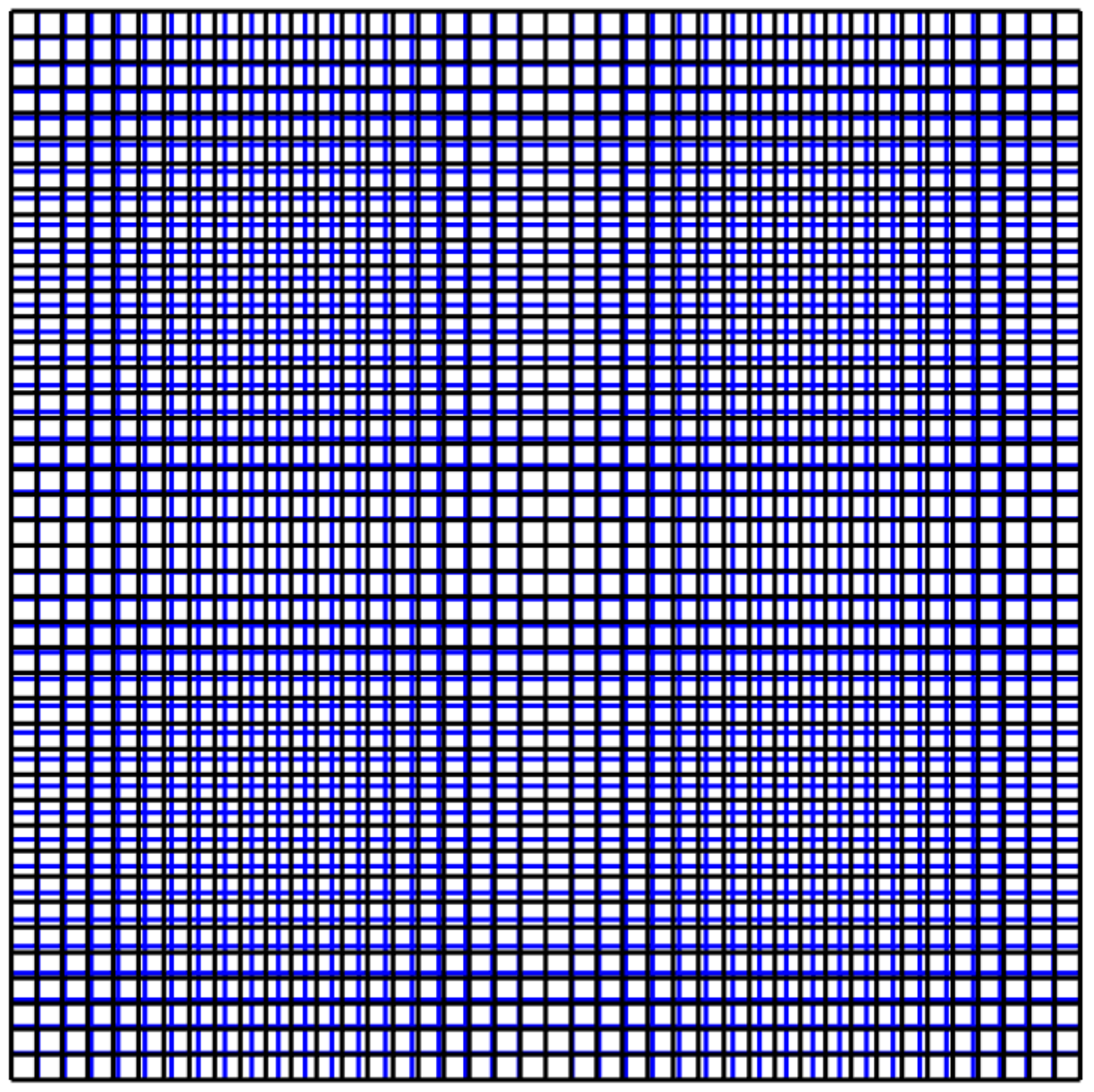}\qquad\qquad\qquad
    \includegraphics[width=.3\linewidth]
                    {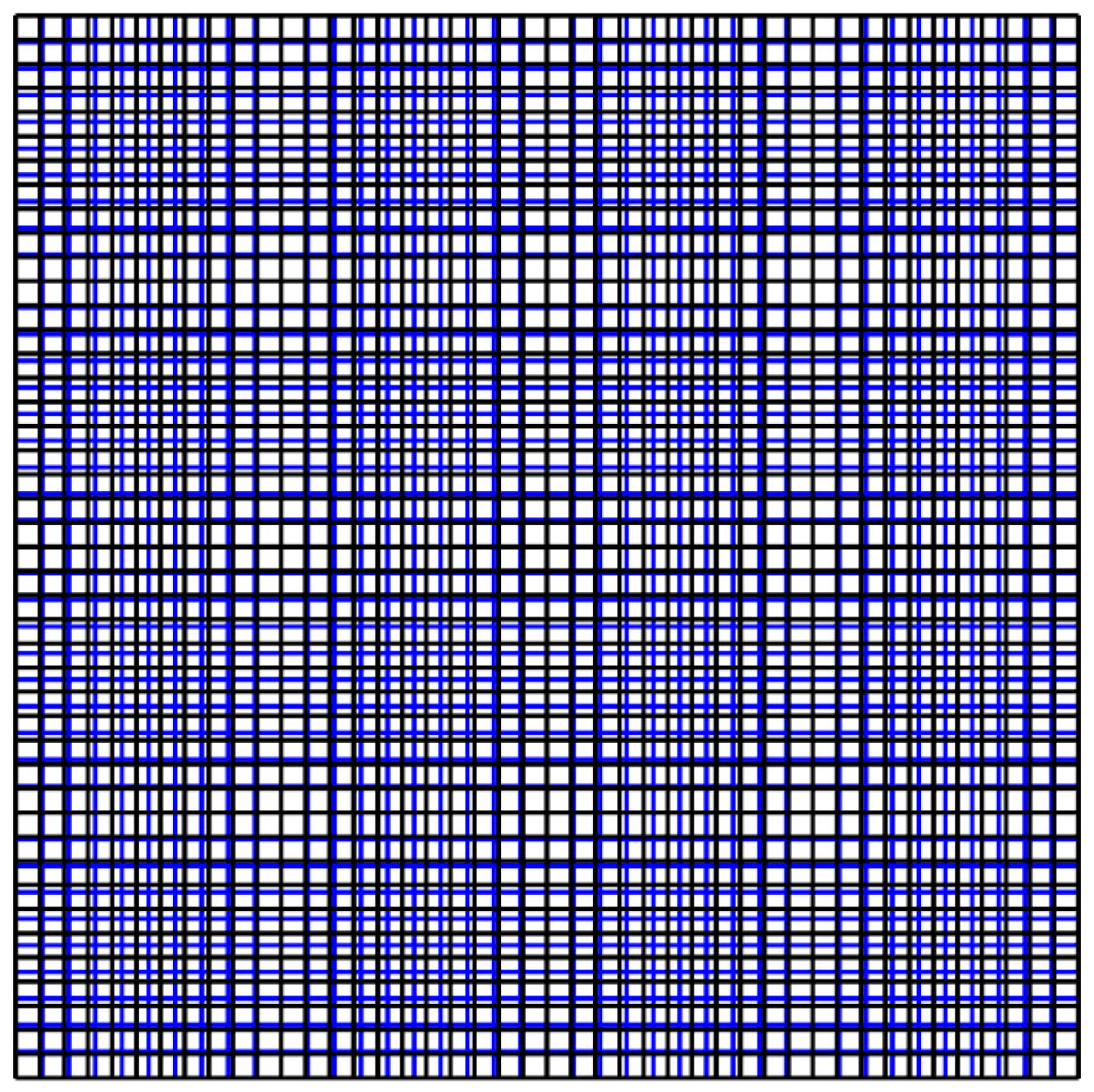}
  \caption{Moir\'e patterns in two parallel square lattices with slightly different lattice constants $h_1$ and $h_2$. Here $h_1/h_2=0.95$ (left) and  $h_1/h_2=0.91$ (right), respectively.  Note that the period of the moir\'e pattern
  decreases as the ratio of lattice constants decreases.}
  \label{moire_fig_sq1}
\end{figure}

\begin{figure}[htb]
\centering
    \includegraphics[width=.3\linewidth]
                    {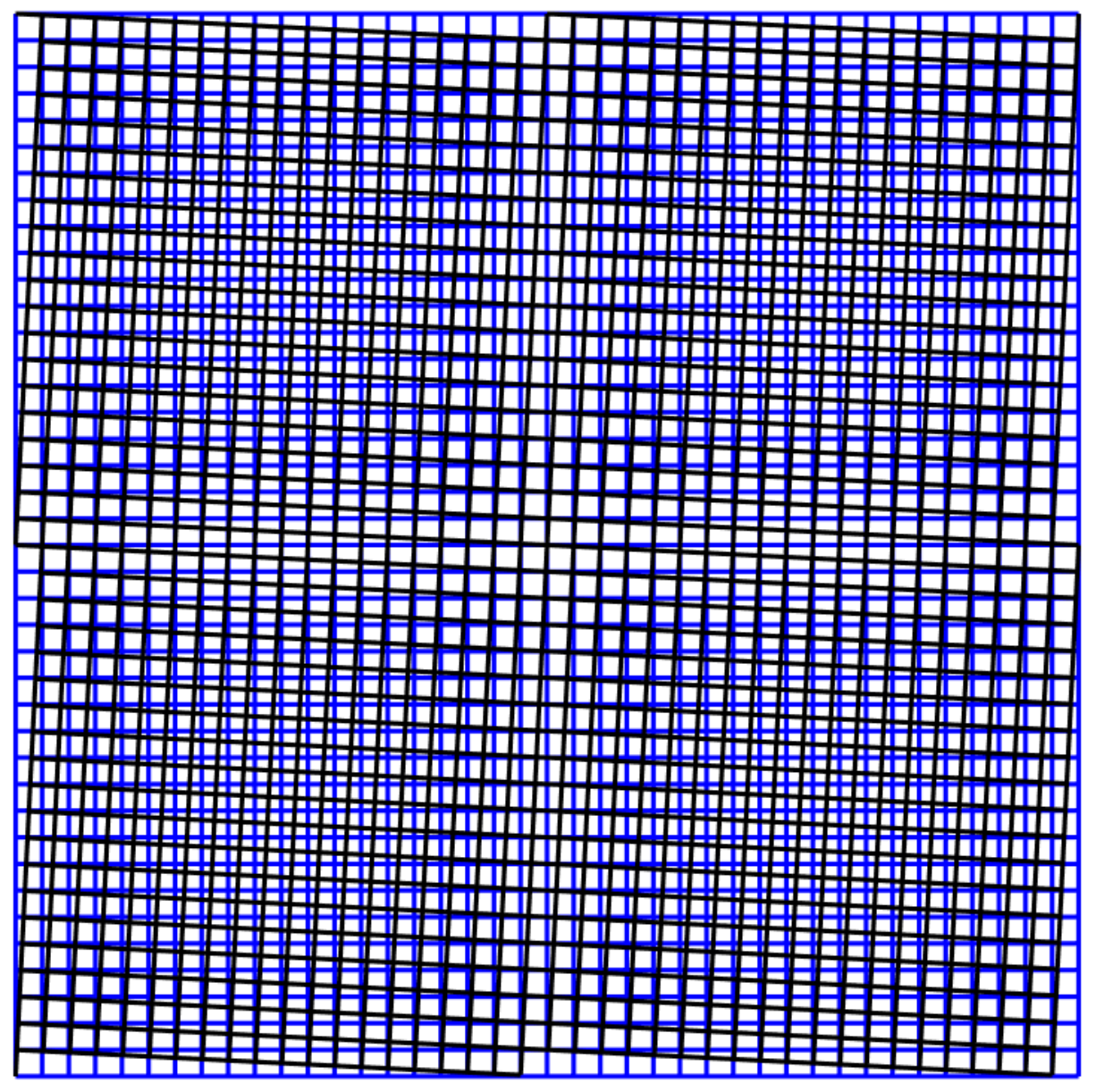}\qquad\qquad\qquad
    \includegraphics[width=.3\linewidth]
                    {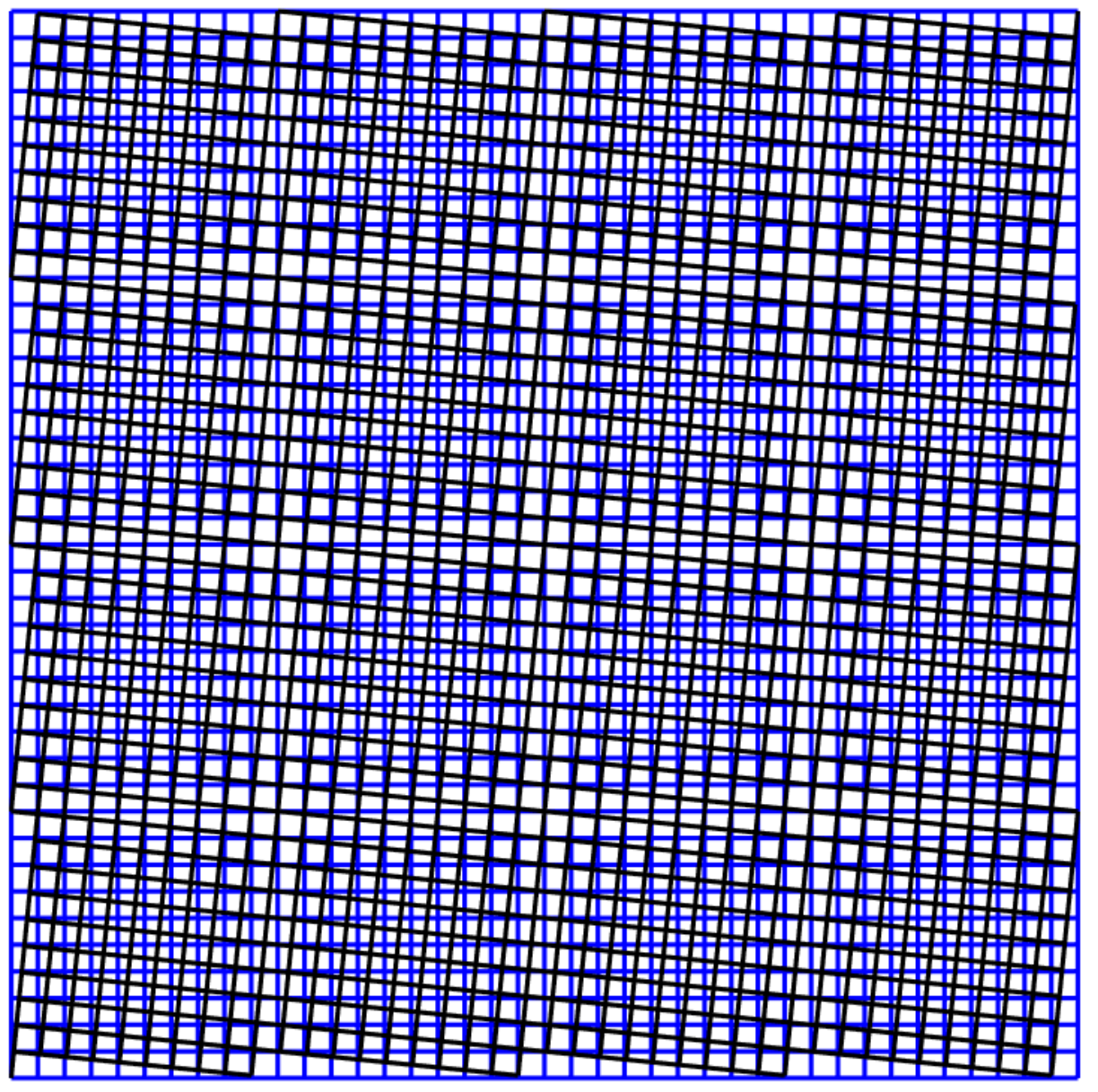}
   \caption{Moir\'e patterns in two identical parallel, and slightly misoriented square lattices with relative rotation of $2.86^\circ$ (left) and $5.71^\circ$ (right), respectively.  Note that the period of the moir\'e pattern
  decreases as the relative rotation increases.}
  \label{moire_fig_sq2}
\end{figure}

\begin{figure}[htb]
\centering
    \includegraphics[width=.3\linewidth]
                    {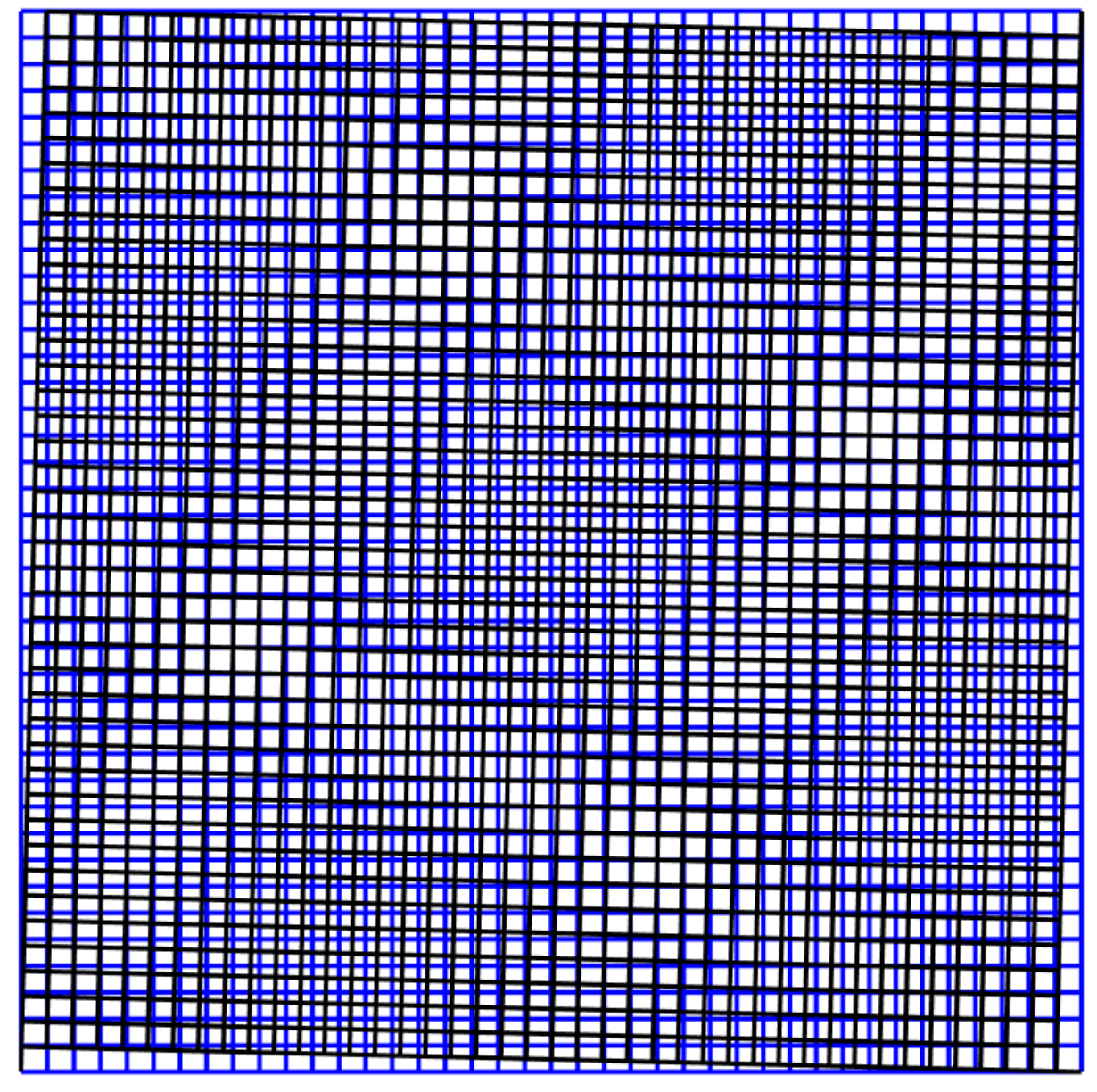}\qquad\qquad\qquad
    \includegraphics[width=.3\linewidth]
                    {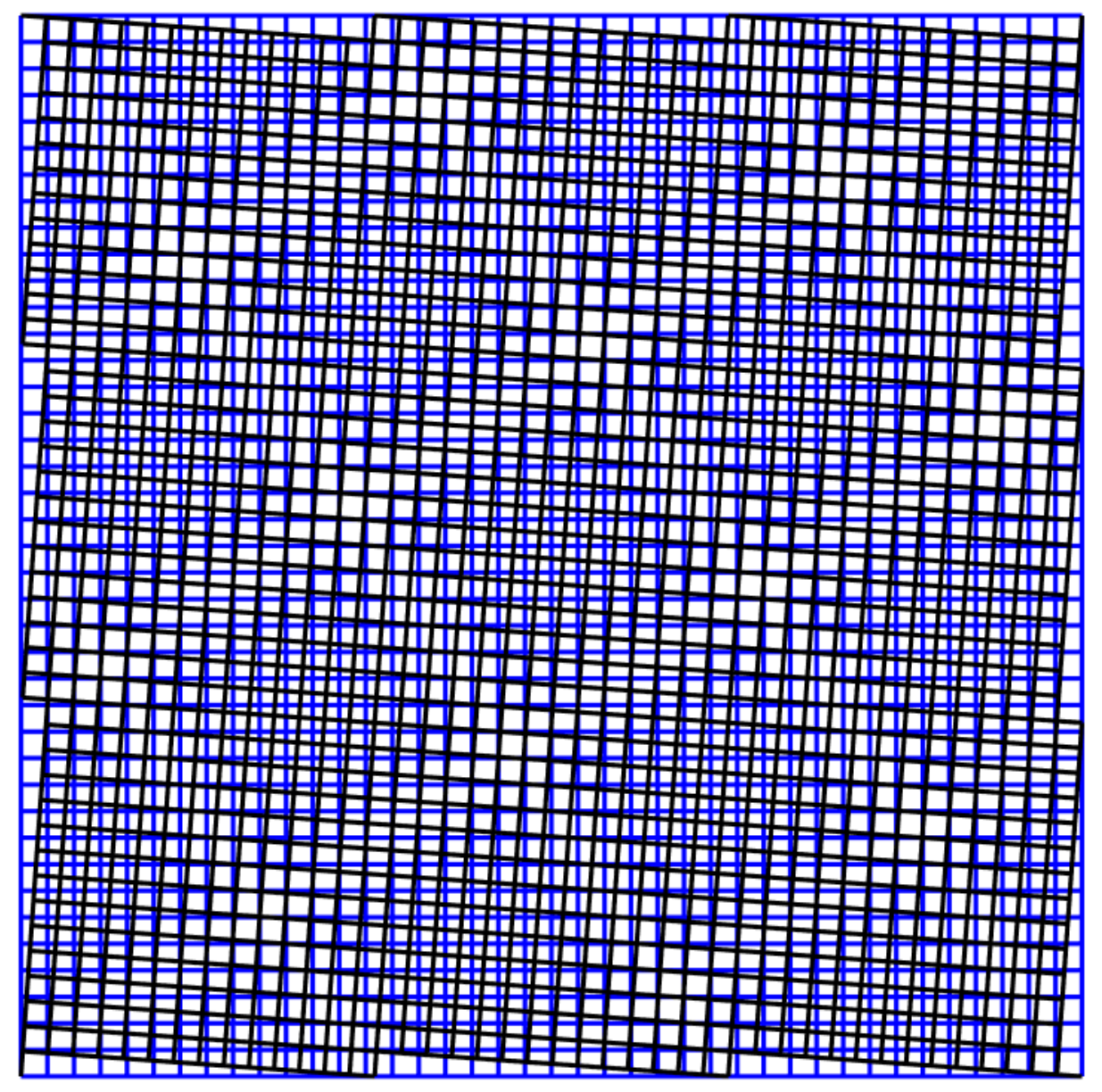}
   \caption{Moir\'e patterns in two slightly misoriented lattices that also have slightly different lattice constants $h_1$ and $h_2$. The relative rotation and the ratio of lattice constants are $1.36^\circ$ and $h_1/h_2=0.95$ (left) and $4.08^\circ$ and $h_1/h_2=0.95$ (right), respectively.}
  \label{moire_fig_sq3}
\end{figure}

One example of this pattern formation in bilayer graphene is called a
relaxed moir\'e pattern \cite{van2015relaxation,van2014moire}.  When two
lattices with different lattice geometries or the same geometry but
different orientations are stacked, a larger periodic pattern, called
a moir\'e pattern, emerges (see Figs.~\ref{moire_fig_hex}-\ref{moire_fig_sq3}). Fig.~\ref{moire_fig_hex} shows an example of the moir\'e pattern in two parallel but slightly rotated identical hexagonal lattices. Notice that the local registry between the lattices varies continuously in space and forms a periodic pattern. In this paper we work with square lattices. Similar moir\'e patterns occur in these lattices because of spatial variations in registry that arise from slightly dissimilar lattice parameters (Fig.~\ref{moire_fig_sq1}) and/or relative rotations (see Figs.~\ref{moire_fig_sq2} and \ref{moire_fig_sq3}).

These moir\'e patterns are a strictly visual
effect.  However, if the atoms on one or both of the lattices are then
relaxed to accommodate the mismatch between the lattices, additional
patterns can occur.  In \cite{van2015relaxation}, the authors study
these relaxed moir\'e patterns by simulating interacting, identical
graphene lattices where one lattice is slightly rotated with respect
to the other.  In \cite{van2014moire}, the authors report on similar
simulations for a slightly rotated graphene lattice interacting with an
h-BN substrate, which also has the structure of a
hexagonal lattice with a slightly larger lattice constant than that of
graphene.  In both papers, simulations in some cases predict a
two-dimensional pattern of domain walls, intersecting at so-called hot spots,
exhibiting large out-of-plane displacements. The domain walls separate
large, flat domains of commensurate regions.

In this paper, we present discrete-to-continuum modeling of a bilayer
of two-dimensional lattices.  Our goal is to develop a continuum model
that can describe how mismatch and misorientation between the lattices
influence the deformation of the bilayer.  For simplicity, we assume that
one of the lattices is rigid and that both lattices are square.  Our discrete-to-continuum procedure yields
a continuum energy with terms describing the elastic energy of the
deformable lattice and a term for the interaction energy between
the deformable and rigid lattices.

For the elastic contributions, our discrete-to-continuum procedure
starts with a square lattice in which the atoms are connected by
extensional, torsional, and dihedral springs that model the resistance
of the lattice to stretching and bending.  The discrete energies of
these springs are upscaled by introducing a small parameter
$\varepsilon$ defined as the ratio of the distance between the
lattices to the lateral extent of the deformable lattice.  Assuming
that the deformed lattice is imbedded in a smooth surface, we expand the
parameterization of this surface and the spring energies in
$\varepsilon$.  Then, an appropriate truncation of the resulting series
yields a continuum elastic energy, which is close to the energy of the
classical von F\"oppl-K\'arm\'an shell theory \cite{PhysRevE.85.066115}.  Our
recent work on a similar one-dimensional problem
\cite{2017arXiv170500072E} indicates that this choice provides a
reasonable generic approximation of the mesoscopic elastic energy.

For the interaction term, the goal of our discrete-to-continuum
procedure is to develop a continuum energy that retains information
about the mismatch between the lattices.  At the discrete level, the
lattices may be mismatched because the rigid and deformable lattices
have different lattice parameters and because of a small rotation
between the lattices.  This mismatch is easily described locally.  The novelty of our model is in
defining an energy density that contains a continuum expression for
the local mismatch.



The continuum energy that we obtain combines the
elastic and weak contributions and has a Ginzburg-Landau-type
structure. The minimizers of the continuum energy represent
equilibrium shapes of the deformable lattice.
To find these equilibrium shapes, we derive the Euler-Lagrange
equations, which are then solved numerically.  We present some basic
comparisons between discrete simulations and the predictions of our
continuum model.  For slightly mismatched layers, our model predicts
large commensurate regions separated by domain walls formed by
localized out-of-plane ridges.  In some cases the number of these
domain walls is determined by the need to accommodate a certain
number of extra rows of atoms on the deformable lattice.  Qualitatively, our
solutions exhibit a pattern of symmetrically spaced hot spots similar
to the predictions of the atomistic simulations in
\cite{van2015relaxation}.



The discrete-to-continuum modeling in this paper generalizes to two
dimensions the work in \cite{2017arXiv170500072E}, in which a
continuum theory for weakly interacting chains of atoms is derived.
The atomistic model includes stretching and bending energies for
strong covalent bonds between atoms in the same chain and an
interaction energy between atoms in adjacent chains.  The
corresponding continuum energy, derived at mesoscopic scale, is of
Ginzburg-Landau type, with an elastic contribution given by the
F\"oppl-von K\'arm\'an energy.  Numerical simulations demonstrate
that the predictions of the continuum model are in close
correspondence with predictions from atomistic simulations.


In \cite{dai2016structure,dai2016twisted}, the authors present a
multiscale model that predicts the deformation of bilayers of graphene
and bilayers of other two-dimensional materials.  In their model, the
total energy of the bilayer has an elastic contribution, associated
with the stretching and bending of the individual layers, and a misfit
energy, which describes the van der Waals interactions between the two
layers.  The misfit energy is defined using the generalized
stacking-fault energy for bilayers, which the authors develop in an
earlier publication \cite{zhou2015van} from density-functional theory
calculations.  The misfit energy is a function of the separation and
disregistry between layers.

The authors use their model to explain the structure of deformed
bilayer graphene in terms of dislocation theory.  In
\cite{dai2016structure}, the model is applied to determine the
structure and energetics of four interlayer dislocations in bilayer
graphene, where the different cases are determined by the angle
between the Burgers vector and the line of dislocation.  In
\cite{dai2016twisted}, the authors use the model to study deformations
that results from a small rotation between the layers.  The model
predicts two distinct equilibrium structures, which the authors call a
breathing mode and a bending mode.  The latter, more stable at small
rotation angles, is characterized by a twist in the dislocation
structure near the dislocation nodes, at which there are also large
out-of-plane displacements.  The authors note that this newly
discovered structure has both different symmetry and period from the
classical moir\'e structure that is often assumed for rotated bilayer
graphene.

The continuum model we develop in this paper has essential elements in
common with the model presented in
\cite{dai2016structure,dai2016twisted}.  Specifically, our model
contains terms for the elastic energy of the deformable layer and a
term for the van der Waals interactions between the two layers.
%
%
However, we derive all terms in our continuum energy by upscaling from
an atomistic description of the problem.  Our upscaling procedure
introduces a small parameter that determines the relative size of the
various contributions to the continuum energy.  Hence we gain insight
into how the balance of these terms produces phenomena like relaxed
moir\'e patterns in interacting bilayers.  Furthermore, our modeling
sets the stage for additional analysis to rigorously determine the
relation between atomistic and continuum descriptions of the problem
\cite{braides2007derivation,braides2014discrete}.


This paper is organized as follows. In Section~\ref{s5}, we formulate a
discrete energy for a system of two weakly interacting square
lattices.  In Section~\ref{s2}, we derive the continuum elastic and
interaction energies. The latter keeps track of the mismatch between
the lattices.  The next section includes numerical results that
compare the atomistic model with the continuum model. Furthermore, in
this section we show how parameters in different ranges give rise to
qualitatively different deformations.  A concluding section
summarizes the paper and is followed by an Appendix containing
computational details of the derivation of the continuum elastic
energy in Section~\ref{s2}.

\section{Atomistic Model} \label{s5}

Suppose that we have a discrete system that consists of two two-dimensional atomic lattices, $\hat{\mathcal A}_1$ and $\hat{\mathcal A}_2$, stacked on top of one another. The atoms on the top lattice $\hat{\mathcal A}_2$ can move and each of these atoms interacts with its neighbors within $\hat{\mathcal A}_2$  via a given strong bond potential. $\hat{\mathcal A}_2$ describes a layer of a two-dimensional material that is nearly inextensible and has a finite resistance to bending. In equilibrium the atoms on $\hat{\mathcal A}_2$ form a square lattice with lattice parameter $h_2$ that occupies a square-shaped, planar domain $D$ with sides of length $L$ and lattice vectors parallel to the sides of $D$.  The atoms on the bottom lattice 
$\hat{\mathcal A}_1$ are fixed at the nodes of another square lattice with lattice constant $h_1$. In this work, $\hat{\mathcal A}_1$ describes a rigid
substrate. All atoms on the lower lattice are assumed to interact with all atoms on the upper lattice via an interatomic
van der Waals potential. In what follows we refer to $\hat{\mathcal A}_1$ as the rigid lattice and to $\hat{\mathcal A}_2$ as the deformable lattice.

\begin{figure}[htb]
\centering
    \includegraphics[width=.5\linewidth]
                    {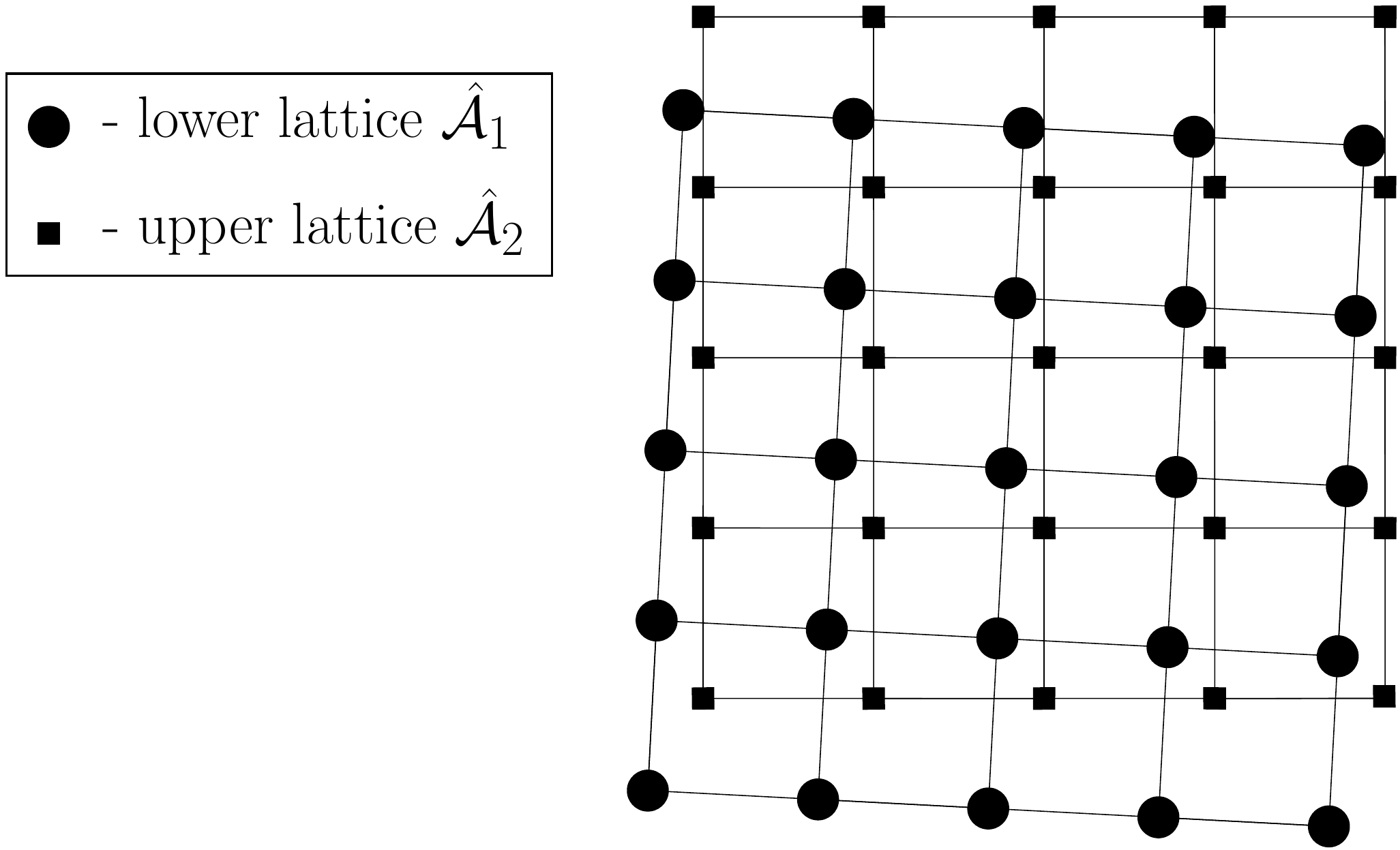}
  \caption{The lattices ${\hat{\mathcal A}}_1$ and ${\hat{\mathcal A}}_2$.}
  \label{f1}
\end{figure}

We assume that, in the reference configuration (Fig.~\ref{f1}), the lattices are
flat and imbedded in two parallel planes, separated by a distance $\sigma$. Here $\sigma$ is equal
to the equilibrium distance between two atoms interacting via the van der Waals potential. 
Note that this reference configuration {\em may not} be in equilibrium. Suppose first that the ratio between $\sigma$ and the equilibrium bond length $h_1$ is large enough (for example, $\sigma/{h_1}>3$ in the case of the Lennard-Jones potential). Then, the van der Waals interactions of an atom on $\hat{\mathcal A}_2$ with atoms on $\hat{\mathcal A}_1$ can be represented by an interaction with the plane with a uniform
atomic density \cite{wilber2007continuum}. In equilibrium, the surfaces  containing $\hat{\mathcal A}_1$ and $\hat{\mathcal A}_2$ are essentially two parallel planes.  The distance between these planes should be slightly smaller than $\sigma$. Indeed, a given atom $a$ on $\hat{\mathcal A}_2$ interacts not only with its closest neighbor $b$ on $\hat{\mathcal A}_1$, but also with the neighbors of $b$ on $\hat{\mathcal A}_1$. If the distance between $a$ and $b$ is $\sigma$, then the forces between $a$ and all neighbors of $b$ are attractive. 

For smaller values of $\sigma/h_1$, the uniform atomic density approximation ignores possible registry
effects that are significant in determining the shape of the deformable lattice $\hat{\mathcal A}_2$. In fact, the only situation in which
the two flat parallel lattices {\it would} correspond to an equilibrium
configuration is when $h_1=h_2$. In this case, all atoms on
$\hat{\mathcal A}_2$ would occupy the positions above the centers of unit
cells formed by the atoms on $\hat{\mathcal A}_1$ and the system would be in {\it global
  registry}. Otherwise, the local equilibrium distance depends on the lattice parameters and the relative orientation of the lattices. Hence, even though the assumed reference configuration is not stress-free, the parameter $\sigma$ is a natural choice for the spacing between $\hat{\mathcal A}_1$ and $\hat{\mathcal A}_2$ in the reference configuration. 

Here we are concerned with the situation when the lattices $\hat{\mathcal A}_1$  and $\hat{\mathcal A}_2$ in the reference configuration have slightly different orientations and/or when $h_1\neq h_2$, but
$|h_1-h_2|/h_1\ll1$ (see Fig.~\ref{f1}). Under these assumptions, global registry cannot
be attained in a flat undeformed configuration. It follows that in order to achieve equilibrium, the
deformable lattice would have to adjust by some combination of bending
and stretching.

Let the current and reference positions of the $N_2^2$ atoms ($N_2:=L/h_2$) on the deformable lattice be given by the set of vectors ${\bf Q}:=\left\{\bq_{ij}\right\}_{i,j=1}^{N_2}\subset \mathbb{R}^{3}$ and ${\bf Q^0}:=\left\{\bq^0_{ij}\right\}_{i,j=1}^{N_2}\subset \mathbb{R}^{3}$, respectively. Because the lattice is periodic, we identify $i=N_2+1$ with $i=1$ and $j=N_2+1$ with $j=1$. In what follows, we use a pair of indices, separated by a comma to denote atoms on $\hat{\mathcal A}_1$ and $\hat{\mathcal A}_2$ but, to avoid clutter, we omit the comma when these indices appear in subscripts. 

For the rigid lattice, the current and the reference configurations are exactly the same. We denote the positions of atoms on the rigid lattice by ${\bf P}:=\left\{{\bf p}_{kl}\right\}_{k,l=-\infty}^{\infty}\subset \mathbb{R}^{3}$. Note that we assume the rigid lattice is infinite in extent in order to appropriately compute the nonlocal van der Waals energy. 

Since the system in the reference configuration consists of two parallel, planar square lattices of atoms, we select an orthonormal basis $\left\{{\bf e}^m_n\right\}_{m,n=1}^2$ for each plane so that the basis vectors are parallel to the respective lattice vectors. Then
\begin{equation}
\label{eq:PQ}
\bp_{kl}=\left\{h_1\left(k\,{\bf e}^1_1+l\,{\bf e}^1_2\right)\right\}_{k,l=-\infty}^{\infty}\subset\mathbb R^2\times\{0\} \mathrm{and}\ \bq^0_{ij}=\left\{h_2\left(i\,{\bf e}_1^2+j\,{\bf e}^2_2\right)\right\}_{i,j={1}}^{N_2}\subset[0,L]^{2}\times\{\sigma\}.
\end{equation}
Here, without loss of generality, we assume that, in the reference
configuration, there is an atom on the deformable lattice $\hat{\mathcal A}_2$ directly above the atom $\hat{\mathcal A}_1$ that lies at the origin.

\begin{figure}[htb]
\centering
    \includegraphics[width=.4\linewidth]
                    {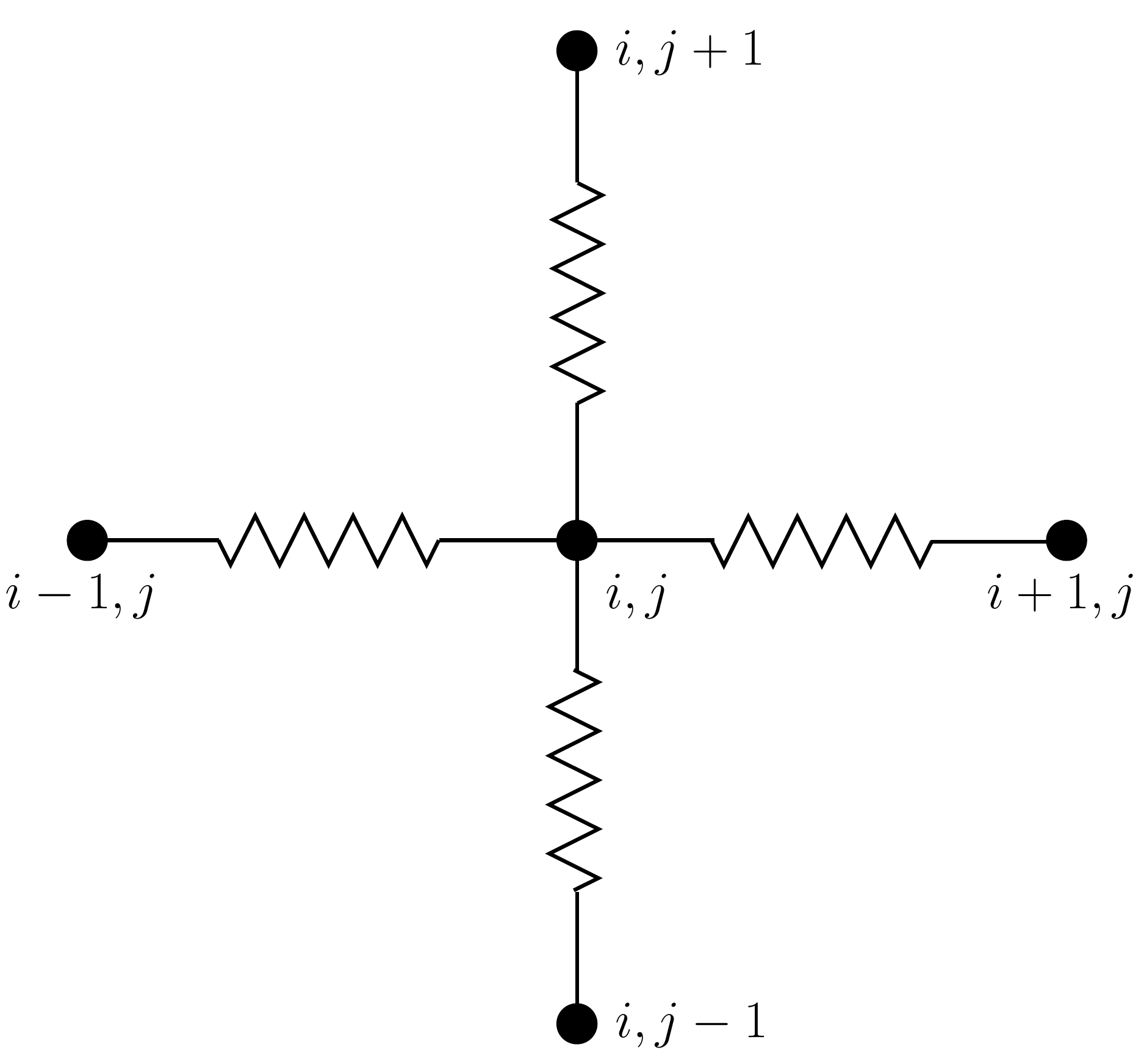}\qquad
    \includegraphics[width=.4\linewidth]
                    {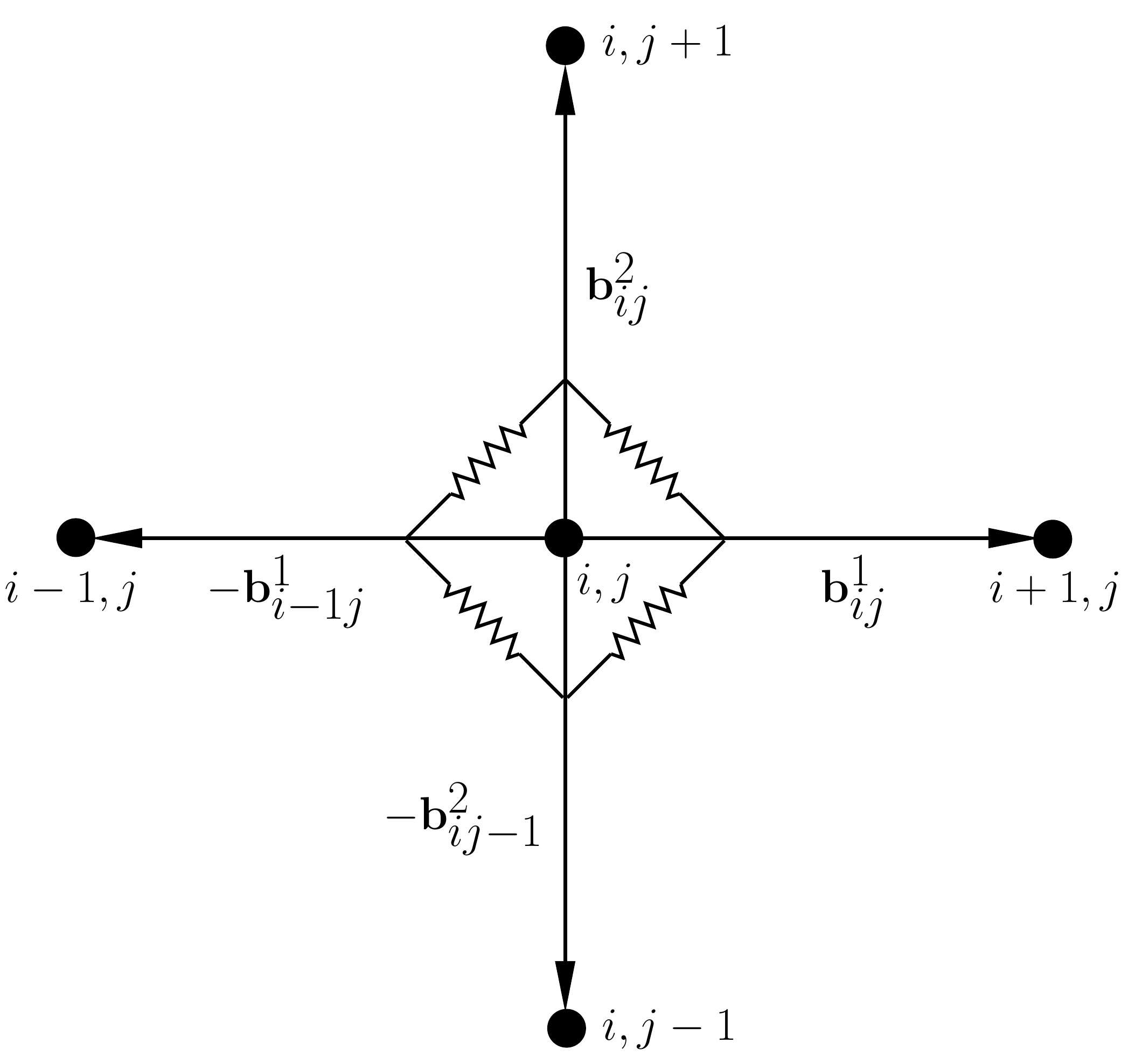}
  \caption{The extensional (left) and torsional (right) springs connected to an atom with indices $i,j$ in ${\hat{\mathcal A}}_2$. Here $i,j=1,\ldots,N_2$.}
  \label{fig:extor}
\end{figure}

For every $i,j=1,\ldots,N_2$, we represent the bonds between the atoms $\bq_{ij}$ and $\bq_{i+1j}$ and the atoms $\bq_{ij}$ and $\bq_{ij+1}$ by the vectors $\bb^1_{ij} = \bq_{i+1j}-\bq_{ij}$ and $\bb^2_{ij} = \bq_{ij+1}-\bq_{ij}$, respectively. We assume that the total energy of the system is given by 
\begin{equation}
E({\bf Q}): = E_s({\bf Q})  + E_t({\bf Q})  + E_d({\bf Q})  + E_w({\bf Q}).
\label{eq:entot}
\end{equation}
Here $E_s$ is the energy required to stretch or compress bonds between two adjacent atoms on $\hat{\mathcal A}_2$, defined by a harmonic potential
\begin{equation}\label{eq:DiscreteE_s}
E_s({\bf Q}) : = \sum_{i,j=1}^{N_2} \frac{k_s}{2}\left[\left( \frac{\|\bb^1_{ij}\|-h_2}{h_2}\right)^2+\left( \frac{\|\bb^2_{ij}\|-h_2}{h_2}\right)^2\right], 
\end{equation}
with $k_s$ being the spring constant. The extensional springs connected to a given atom with the indices $i,j$ are shown in Fig.~\ref{fig:extor} (left); note that due to periodicity of the lattice only two of these springs per atom appear in the sum in \eqref{eq:DiscreteE_s}.

The bending between the interatomic bonds is penalized by introducing harmonic torsional and dihedral springs between the bonds. The bending energy associated with the torsional springs is given by
\begin{multline}
  E_{t}({\bf Q})
  :=
  \sum_{i,j=1}^{N_2} \frac{k_t}{2}\left[\left(\theta\left(\bb^1_{ij},\bb^2_{ij}\right)-\pi/2\right)^2+\left(\theta\left(\bb^2_{ij},-\bb^1_{i-1j}\right)-\pi/2\right)^2\right.\\ \left. +\left(\theta\left(-\bb^1_{i-1j},-\bb^2_{ij-1}\right)-\pi/2\right)^2+\left(\theta\left(-\bb^2_{ij-1},\bb^1_{ij}\right)-\pi/2\right)^2\right], \label{e21}
\end{multline}
where $k_t$ is the torsional spring constant and $\theta({\bf a},{\bf c})$ is the angle between the vectors ${\bf a}$ and ${\bf c}$. The configuration of the torsional springs associated with a given atom in $\hat{\mathcal A}_2$ is shown in Fig.~\ref{fig:extor} (right). Assuming that admissible in-plane deformations of $\hat{\mathcal A}_2$ are small, in the sequel we consider the expression 
\begin{equation}
  E_{t}({\bf Q})
  =
  \sum_{i,j=1}^{N_2} \frac{k_t}{2}\left[\frac{{\left(\bb^1_{ij}\cdot\bb^2_{ij}\right)}^2}{{\left\|\bb^1_{ij}\right\|}^2{\left\|\bb^2_{ij}\right\|}^2}+\frac{{\left(\bb^2_{ij}\cdot\bb^1_{i-1j}\right)}^2}{{\left\|\bb^2_{ij}\right\|}^2{\left\|\bb^1_{i-1j}\right\|}^2}+\frac{{\left(\bb^1_{i-1j}\cdot\bb^2_{ij-1}\right)}^2}{{\left\|\bb^1_{i-1j}\right\|}^2{\left\|\bb^2_{ij-1}\right\|}^2}+\frac{{\left(\bb^2_{ij-1}\cdot\bb^1_{ij}\right)}^2}{{\left\|\bb^2_{ij-1}\right\|}^2{\left\|\bb^1_{ij}\right\|}^2}\right]
\label{e21.5}
\end{equation}
 for the bending energy, which is equivalent to \eqref{e21} to leading order. 
 
 The expressions for the extensional and torsional springs, respectively, show that the sum of the corresponding energy components is minimized when all unit cells of the lattice $\hat{\mathcal A}_2$ are squares with the side of the length $h_2$. Note, however, that the lattice can be folded along the directions parallel to the sides of the domain $D$ without incurring any energy cost. The appropriate cost can be added by incorporating dihedral springs into the lattice.

\begin{figure}[htb]
\centering
    \includegraphics[width=.4\linewidth]
                    {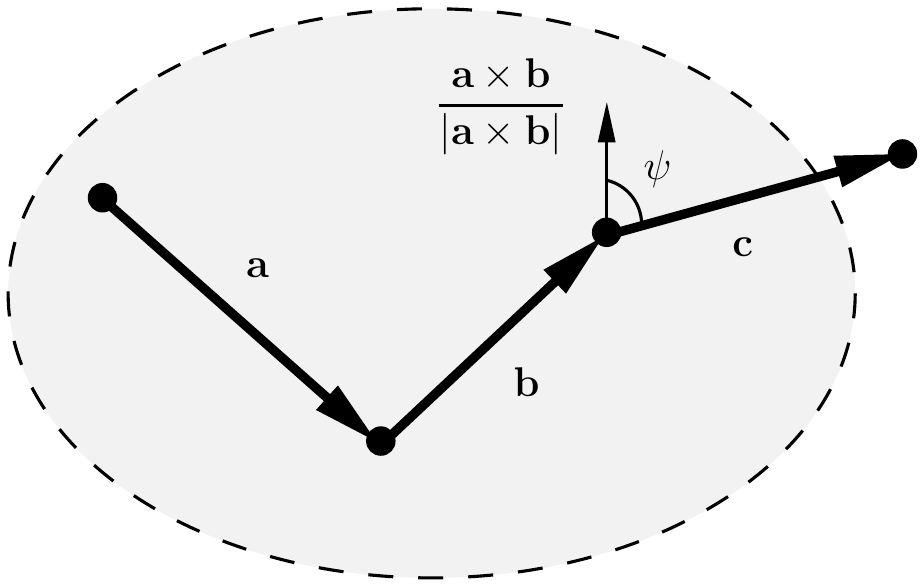}
  \caption{A dihedral spring connecting vectors ${\bf a,\ b,}\ \mathrm{and}\ {\bf c}$. The spring energy is minimized when $\psi=\frac{\pi}{2}$.}
  \label{fig:dih}
\end{figure}

\begin{figure}[htb]
\centering
    \includegraphics[width=.3\linewidth]
                    {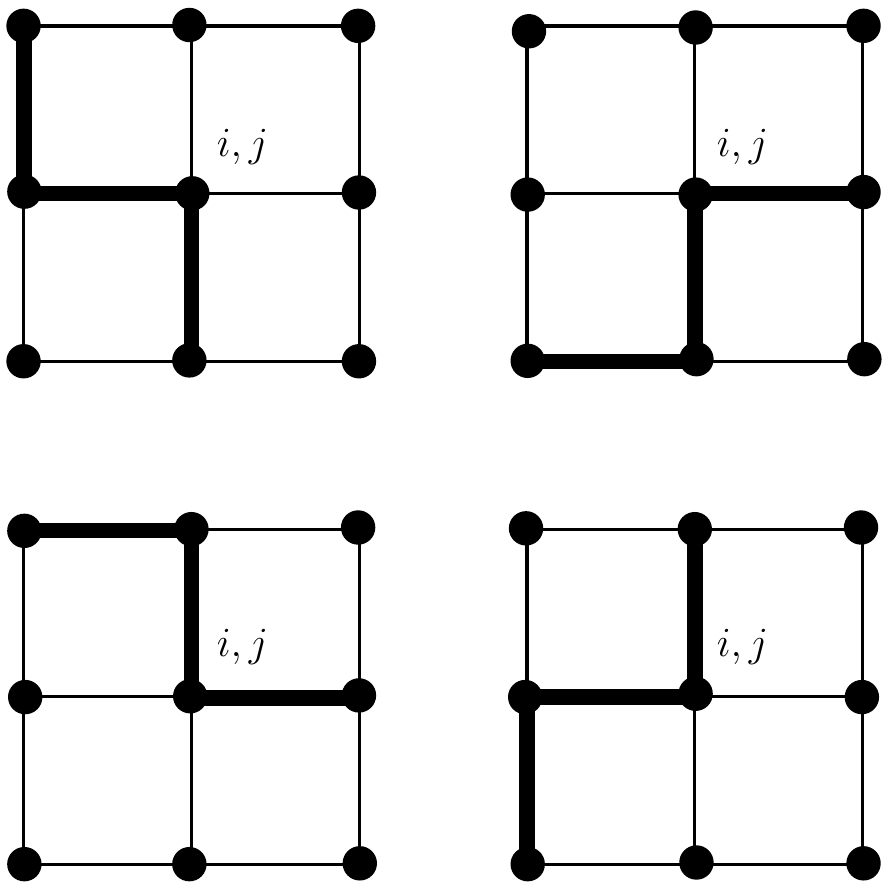}
  \caption{The dihedral springs connecting the atom $i,j$ to its neighbors in $\hat{\mathcal A}_2$. Here $i,j=1,\ldots,N_2$.}
  \label{fig:dih_con}
\end{figure}

A dihedral spring connects three adjacent bonds so that this spring energy is minimized when the third bond lies in the plane formed by the first two bonds (Fig.~\ref{fig:dih}).
A prototypical dihedral spring energy that satisfies this condition is given by the expression 
\[e_d({\bf a,b,c}):=\frac{k_d}{2}\cos^2\psi=\frac{k_d}{2}\frac{{\left(({\bf a}\times{\bf b})\cdot{\bf c}\right)}^2}{{\|{\bf a}\times{\bf b}\|}^2{\|{\bf c}\|}^2},\]
where $k_d$ is the dihedral spring constant and $\psi$ is the dihedral angle defined as in Fig.~\ref{fig:dih}. We assume that, for each $i,j=1,\ldots,N_2$, the atom $\bq_{ij}$  in the lattice $\hat{\mathcal A}_2$ is connected to all dihedral springs shown in Fig.~\ref{fig:dih_con}. Note that the actual number of the dihedral springs connected to this and all other atoms is larger due to periodicity of the lattice. 
The bending energy associated with the dihedral springs is then given by
\begin{multline}
  E_{d}({\bf Q})
  :=
  \sum_{i,j=1}^{N_2} \frac{k_d}{2}
  \left[
    \frac{{\left((\bb^1_{i-1j}\times\bb^2_{ij-1})\cdot\bb^2_{i-1j}\right)}^2}{{\|\bb^1_{i-1j}\times\bb^2_{ij-1}\|}^2{\|\bb^2_{i-1j}\|}^2}+
    \frac{{\left((\bb^1_{ij}\times\bb^2_{ij-1})\cdot\bb^1_{i-1j-1}\right)}^2}{{\|\bb^1_{ij}\times\bb^2_{ij-1}\|}^2{\|\bb^1_{i-1j-1}\|}^2}\right.\\ \left.
    +\frac{{\left((\bb^1_{ij}\times\bb^2_{ij})\cdot\bb^1_{i-1j+1}\right)}^2}{{\|\bb^1_{ij}\times\bb^2_{ij}\|}^2{\|\bb^1_{i-1j+1}\|}^2}
    +\frac{{\left((\bb^2_{ij}\times\bb^1_{i-1j})\cdot\bb^2_{i-1j-1}\right)}^2}{{\|\bb^2_{ij}\times\bb^1_{i-1j}\|}^2{\|\bb^2_{i-1j-1}\|}^2}\right]. \label{e_dih}
\end{multline}

The energy of the weak van der Waals interaction between $\hat{\mathcal A}_1$ and $\hat{\mathcal A}_2$ is given by
\begin{equation}\label{eq:DiscreteE_w}
  E_w({\bf Q}) 
  =
  \omega\sum_{i,j=1}^{N_2} \sum_{m,n=-\infty}^{\infty}  g\left(\frac{\|\bq_{ij}-\bp_{mn} \|}{\sigma}\right),
\end{equation}  
where $g$ is a given weak pairwise interaction potential.  The parameters $\sigma$ and $\omega$ define the equilibrium interatomic distance and the strength of the potential energy \eqref{eq:DiscreteE_w}, respectively. In what follows, we assume that $g$ is the classical Lennard-Jones 12-6 potential given by
\begin{equation}
g(r) = r^{-12}-2r^{-6}.
\label{ede29}
\end{equation} 
Note that the inner sum in \eqref{eq:DiscreteE_w} is taken over the entire rigid lattice to properly account for weak interactions between the lattices.

\section{Continuum Model} \label{s2}

Next, we briefly describe the approach we take to derive the continuum model. We assume that the atoms of the deformable lattice $\hat{\mathcal A}_2$ are imbedded in a smooth surface $\mathcal A_2\subset\mathbb R^3$ and describe this surface parametrically in terms of the displacement field. Nondimensionalizing the discrete problem introduces a small geometric parameter $\varepsilon=\sigma/L$, equal to the ratio of the equilibrium van der Waals distance to the length of the side of the domain $D$. Evaluating the displacements at atomic positions, substituting these into the expression \eqref{eq:entot} for the discrete energy, expanding the result in terms of $\varepsilon$, and converting summation into integration, leads to an expansion in terms of $\varepsilon$ for the continuum energy, written as a functional of the displacement field.

We identify the leading order terms in this expansion, up to the order at which contributions from the extensional, torsional, and dihedral springs, as well the van der Waals interactions are included. The resulting continuum energy is of Ginzburg-Landau type and contains terms of different powers in $\varepsilon$. The minimizers of the continuum energy typically exhibit bulk regions of registry, separated by thin walls where the gradient of the displacement field is large. Thus, within the walls, the contributions from higher order terms generally cannot be neglected. We choose to cut off the expansion that leads to the continuum energy at the order when all components of the displacement contribute to the energy density inside the walls at leading order. Finally, in the next section, we present the results of simulations confirming that the behavior of minimizers of the continuum energy match that of minimizers of the discrete energy.

Both the discrete and continuum nondimensional models contain the small parameter $\varepsilon$ and, in particular, the continuum model cannot be thought of as a limit of the discrete model as $\varepsilon\to0$. Instead, we conjecture that both models converge to the same asymptotic limit as $\varepsilon\to0$ in the appropriate sense. The limit has to be understood within the framework of $\Gamma$-convergence \cite{braides2002gamma} so that both energies are $\Gamma$-{\it equivalent} \cite{braides2008asymptotic}. Hence the number of terms retained in the expansion of the discrete problem in order to obtain the continuum problem should be sufficient to reproduce the behavior of the discrete system for a small $\varepsilon$. The proof of $\Gamma$-equivalence is a subject of a future work.

As noted above, a first step in formally deriving a continuum model is to assume that the deformed lattice $\hat{\mathcal
A}_2$ is embedded in a sufficiently smooth surface $\mathcal A_2$.  We denote this surface in the reference configuration by $\mathcal A^{0}_2$ and set
\begin{equation}
  \mathcal A^{0}_2 = \left\{(x_1,x_2,\sigma):x_1,x_2\in[0,L]\right\}.
  \label{ee14}
\end{equation}

We let $(u_1(x_1,x_2),u_2(x_1,x_2),v(x_1,x_2))$ be the displacement of the point $(x_1,x_2,\sigma)$
on $\mathcal A^{0}_2$.  Hence the deformed
surface $\mathcal A_2$ is given by
\begin{equation}
  \left\{(x_1+u_1(x_1,x_2),x_2+u_2(x_1,x_2),\sigma+v(x_1,x_2)):x_1,x_2\in[0,L]\right\}. 	
  \label{ee12}
\end{equation}
In particular, an atom at the point $(x_{1}^{ij},x_{2}^{ij},\sigma)$
on $\mathcal A^{0}_2$ is displaced to the point
$(x_{1}^{ij}+u_1(x_{1}^{ij},x_{2}^{ij}),x_{2}^{ij}+u_2(x_{1}^{ij},x_{2}^{ii}),\sigma+v(x_{1}^{ij},x_{2}^{ij}))$. 

We assume $\sigma<<L$, i.e., that the spacing between the planes is much
less than the lateral extent of the system.  To exploit this, we set ${\bf x}=(x_1,x_2)$, ${\bf u}=(u_1,u_2)$ and rescale as follows
\begin{equation}
   \boldsymbol\chi = \frac{\bf x}{L},\ \ 
  \boldsymbol\xi = \frac{\bf u}{\varepsilon L},\ \ 
  \eta = \frac{v}{\varepsilon L}, \ \
  \mathcal E=\frac{\varepsilon E}{\omega} .
  \label{ee1}
\end{equation}
This gives the nondimensional parameters
\begin{equation}
  \varepsilon = \frac{\sigma}{L},\ \ 
  \delta_{1} = \frac{h_{1}}{\sigma},\ \ 
  \delta_{2} = \frac{h_{2}}{\sigma},\ \ 
  \gamma_s=\frac{k_s}{\omega\delta_2^2}, \ \
  \gamma_t=\frac{8k_t}{\omega\delta_2^2}, \ \
  \gamma_d=\frac{2k_d}{\omega}.
  \label{ee1.1}
\end{equation}
The scalings for the displacements are appropriate for small deformations considered here and
eventually lead to expressions for the strains similar to those for F\"oppl-von
K\'arm\'an theory.  The constants in the definitions of $\gamma_t$ and $\gamma_d$ appear to simplify the expressions for the continuum elastic energy.

We obtain with a slight abuse of notation that
\begin{equation}
  \mathcal A_2
  =
  \left\{(\chi_1+\varepsilon \xi_1(\boldsymbol\chi),\chi_2+\varepsilon \xi_2(\boldsymbol\chi),\varepsilon+\varepsilon \eta(\boldsymbol\chi))\colon \boldsymbol\chi\in[0,1]^2\right\}
  =
  \left\{(\boldsymbol\chi+\varepsilon {\boldsymbol\xi}(\boldsymbol\chi),\varepsilon+\varepsilon \eta(\boldsymbol\chi))\colon \boldsymbol\chi\in[0,1]^2\right\}.
  \label{ee13}
\end{equation}
We assume that $\delta_i=\mathcal{O}(1),\ i=1,2$,
i.e., the lattice parameters for $\hat{\mathcal A}_1$ and $\hat{\mathcal A}_2$ are comparable to the distance
between $\hat{\mathcal A}_1$ and $\hat{\mathcal A}_2$ (and hence both are much smaller than the lateral extent of the
system).  Furthermore, in order to observe the registry effects on a macroscale,
we assume that
\begin{equation}
  \frac{\delta_{1}-\delta_{2}}{\varepsilon\delta_2}:=\alpha=\mathcal{O}(1)
  \label{ee9},
\end{equation}
so that the mismatch between the equilibrium lattice parameters of $\hat{\mathcal A}_1$ and $\hat{\mathcal A}_2$ is small.

In the rescaled coordinates, the atoms on $\mathcal{A}_{2}^0$ are located at the points $\bq_{ij}^0=(\boldsymbol\chi_{ij},\eps)$, where $\boldsymbol\chi_{ij}=\eps\delta_2\left(i{\bf e}_1^2+j{\bf e}_2^2\right)$ for $i,j=1,\ldots,N_2$ are obtained by dividing ${\bf q}_{ij}^0$ by $L$ in \eqref{eq:PQ}. The $i,j$-th atom is then displaced to the point
\begin{equation}
\bq_{ij}=(\boldsymbol\chi_{ij}+\eps\boldsymbol\xi(\boldsymbol\chi_{ij}),\eps+\eps\eta(\boldsymbol\chi_{ij})),
\label{qij}
\end{equation}
for every $i,j=1,\ldots,N_2$. Note that here and in what follows we continue to use the notation ${\bf q}_{ij}$, ${\bf q}_{ij}^0$, and ${\bf b}_{ij}^k$, but now to denote the corresponding nondimensional quantities.

\subsection{Elastic Energy Contribution} \label{elencon}

Our treatment of the discrete-to-continuum limit for the elastic energy is consistent with a number of recent studies \cite{le2017hexagonal}-\nocite{blanc_lebris_lions_2007,S0218202512500327,Schmidt2008}\cite{Blanc2002}. The principal idea is to exploit the smallness of the parameter $\eps$. By using the Taylor expansions of $\boldsymbol\xi(\boldsymbol\chi_{ij})$ and $\eta(\boldsymbol\chi_{ij})$ in $\eps$, each bond ${\bf b}_{ij}^k$ can be written as an asymptotic series in $\varepsilon$. Substituting the appropriate expansions into the expressions \eqref{eq:DiscreteE_s}, \eqref{e21.5}, and \eqref{e_dih} for the extensional, torsional, and dihedral energy components, respectively, and taking into account the energy rescaling in \eqref{ee1}, we can redefine both energies in terms of values of $\boldsymbol\xi$ and $\eta$ at $\boldsymbol\chi_{ij}$, where $i,j=1,\ldots,N_2$. We have that
\begin{multline}
\label{stretch}
  \mathcal E_s[\boldsymbol\xi,\eta]=
  \sum_{i,j=1}^{N_2} \frac{k_s}{2\omega}
  \left[
    (\xi_{1,1}^{2}+\xi_{2,2}^{2})\varepsilon^{3}
    +
    \left(
    \delta_{2}(\xi_{1,1}\xi_{1,11} + \xi_{2,2}\xi_{2,22}) \right.\right. \\
    \left.\left.+\,
    \eta_{,1}^{2}\xi_{1,1} + \eta_{,2}^{2}\xi_{2,2}
    +
    \xi_{1,1}\xi_{2,1}^{2} + \xi_{2,2}\xi_{1,2}^{2}
  \right)
  \varepsilon^{4}
  \right]
  +
  \mathcal{O}(\varepsilon^{5}),
  \end{multline}
while
\begin{equation}
\label{bend}
   \mathcal E_t[\boldsymbol\xi,\eta]
  =
  \sum_{i,j=1}^{N_2} \frac{k_t}{\omega}
  \left(
    2(\xi_{1,2}+\xi_{2,1})^{2}\varepsilon^{3}
    +
    4(\xi_{1,2}+\xi_{2,1})
      \left(
      -
      \nabla \xi_{1}\cdot \nabla \xi_{2}
      +
      \eta_{,1}\eta_{,2}
      \right)\varepsilon^{4}
  \right)
   +  
  \mathcal{O}(\varepsilon^{5}),
\end{equation}
and 
\begin{equation}
   \mathcal E_d[\boldsymbol\xi,\eta]
  =
  \sum_{i,j=1}^{N_2} \frac{k_d\delta_{2}^{2}}{\omega}
  \left[
    \eta_{,11}^2+2\eta_{,12}^2+\eta_{,22}^2
    \right]\varepsilon^{5}
  +
  \mathcal{O}(\varepsilon^{6}).  
  \label{dihe}
\end{equation}
The details of derivations that led to these expansions are given in the Appendix. 

Per the discussion above, we would like to truncate these energy expansions in such a way that the limiting behavior of the minimizers of the truncated energy is in some way close to the behavior of minimizers of the original discrete model. Here we impose the following three criteria on the truncated model: (i) it should be well-posed mathematically; (ii) it should preserve all relevant interactions between atoms; and (iii) it should respect the standard invariance assumptions of continuum mechanics.

The first criterion limits the choice of where to terminate the expansions. For example, there could be sign constraints placed on highest derivative terms to guarantee that the continuum variational problem has a minimum. However, for some truncations of the energy expansion, these constraints might not be satisfied and the variational problem is not solvable. Adding the additional, higher derivatives terms would typically yield a model that can be solved. The downside of this, however, is that the model quickly becomes extremely complicated. 

Because we expect the minimizers of the discrete energy \eqref{eq:entot} to develop domain walls of characteristic width $\eps$, we should also expect some derivatives of the minimizers to appear as powers of $\eps^{-1}$ inside the walls. Accordingly, all terms in the expansions \eqref{stretch}-\eqref{dihe} may then contribute roughly the same amount to the overall energy, making the asymptotic procedure that led to \eqref{stretch}-\eqref{dihe} invalid. Note that as an alternative to formulating a purely continuum theory, here one might opt to use a quasicontinuum method, in which the singular regions are resolved using the original discrete formulation \cite{blanc_lebris_lions_2007}.

On the other hand, in the regions between the walls, all minimizers should have bounded derivatives and the terms with higher powers of $\eps$ in \eqref{stretch}-\eqref{dihe} should simply provide small corrections to the lower order contributions. The situation is not unlike that arising in continuum modeling of crystalline solids, where the structural defects---such as dislocations---are described only in terms of their influence on the global strain field, without properly resolving the defect core. 

Motivated by these ideas, we make the following choices when truncating $\mathcal E_s,\ \mathcal E_t,$ and $\mathcal E_d$:

\begin{equation}
\label{stretchs}
  \mathcal E_s[\boldsymbol\xi,\eta]\sim
  \sum_{i,j=1}^{N_2} \frac{k_s\varepsilon^{3}}{2\omega}
  \left[
    {\left(\xi_{1,1}+ \frac{\varepsilon \eta_{,1}^{2}}{2}\right)}^2+{\left(\xi_{2,2}+\frac{  \varepsilon\eta_{,2}^2}{2}\right)}^2
  \right],
  \end{equation}
\begin{equation}
\label{bends}
   \mathcal E_t[\boldsymbol\xi,\eta]
  \sim
  \sum_{i,j=1}^{N_2} \frac{8\,k_t\varepsilon^{3}}{\omega}
  {\left[
    \frac{\xi_{1,2}+\xi_{2,1}}{2}
    +
    \frac{\varepsilon}{2}\eta_{,1}\eta_{,2}
  \right]}^2,
\end{equation}
and 
\begin{equation}
   \mathcal E_d[\boldsymbol\xi,\eta]
  \sim
  \sum_{i,j=1}^{N_2} \frac{k_d\delta_{2}^{2}\varepsilon^{5}}{\omega}
  \left[
    \eta_{,11}^2+2\eta_{,12}^2+\eta_{,22}^2
    \right].  
  \label{dihes}
\end{equation}
Note that here we neglected some third-order terms in $\varepsilon$, in particular, the terms that contain second-order derivatives in $\boldsymbol\xi$ or are cubic in derivatives of $\boldsymbol\xi$. Further, we incorporated some quartic terms in the derivative of $\eta$, that allow us to complete squares in \eqref{stretchs} and \eqref{bends}. We conjecture that incorporating/deleting these higher order terms from the truncated energy still gives minimizers with the structure close to that of the minimizers of the discrete energy as $\varepsilon\to0$.

We now recall that $\mathcal A_2^0$ has a unit area in nondimensional coordinates and that the spacing between the atoms is equal to $\eps\delta_2\ll1$. Hence, the number of atoms on $\mathcal A_2^0$ is $\sim\frac1{\eps^2}$ and therefore 
\[
\mathcal E_s[\boldsymbol\xi,\eta]\sim\frac{\gamma_s\eps}{2}\int_{[0,1]^2} \left[
    {\left(\xi_{1,1}+ \frac{\varepsilon \eta_{,1}^{2}}{2}\right)}^2+{\left(\xi_{2,2}+\frac{  \varepsilon\eta_{,2}^2}{2}\right)}^2
  \right]\,d{\boldsymbol\chi}=:\mathcal F_s^\eps[\boldsymbol\xi,\eta]
\]
while
\[
\mathcal E_t[\boldsymbol\xi,\eta]\sim\gamma_t\eps\int_{[0,1]^2} {\left[
    \frac{\xi_{1,2}+\xi_{2,1}}{2}
    +
    \frac{\varepsilon}{2}\eta_{,1}\eta_{,2}
  \right]}^2\,d{\boldsymbol\chi}=:\mathcal F_t^\eps[\boldsymbol\xi,\eta]
\]
and
\begin{equation}
\label{bend_cont}
\mathcal E_d[\boldsymbol\xi,\eta]\sim\frac{\gamma_d\eps^3}{2}\int_{[0,1]^2} \left[
    \eta_{,11}^2+2\eta_{,12}^2+\eta_{,22}^2
    \right]\,d{\boldsymbol\chi}=:\mathcal F_d^\eps[\boldsymbol\xi,\eta].
\end{equation}

 \subsection{Van der Waals Energy Contribution} \label{vaencon}

We now derive the continuum version of \eqref{eq:DiscreteE_w},
which is the contribution to the energy from the van der Waals
interactions.  We shall see that the continuum version has the form
\begin{equation}
  \mathcal F^\eps_w[\boldsymbol\xi,\eta]
  =
  \frac{1}{\varepsilon}\int_{[0,1]^2} G\left(\boldsymbol\chi,\boldsymbol\xi,\eta\right)\,d\boldsymbol\chi.
  \label{vdwfin}
\end{equation}
The novelty of our model is in defining a function $G$ that gives a
continuum description of the lattice mismatch that arises from incommensurability.  We shall first focus in the inner double sum on the
right-hand side of \eqref{eq:DiscreteE_w} and try to estimate the
interaction of a given atom on the deformable lattice with all the
atoms on the rigid lattice.  We accomplish this by developing an
expression for the local mismatch
between the two lattices
as a function of $\boldsymbol\chi,\ \boldsymbol\xi$, and $\eta$.

Our starting point is to pick an atom $i,j$ on $\hat{\mathcal A}_2$.  A
discrete description of the total interaction energy between this atom
and the atoms on the rigid lattice $\hat{\mathcal A}_1$ is given by
\begin{equation}
  \sum_{m,n=-\infty}^{\infty}  g\left( d_{ij}^{mn}/\varepsilon\right),
  \label{ee3}
\end{equation}
where $g$ is defined in \eqref{ede29} and $d_{ij}^{mn}$ is the distance between the fixed atom $i,j$ on
$\hat{\mathcal{A}}_{2}$ and the atom $m,n$ on $\hat{\mathcal{A}}_{1}$. Note that here we could use a finite sum instead due to the fast decay of the interaction potential $g$ with distance. 
%


%



\begin{figure}[htb]
\centering
    \includegraphics[width=.6\linewidth]
                    {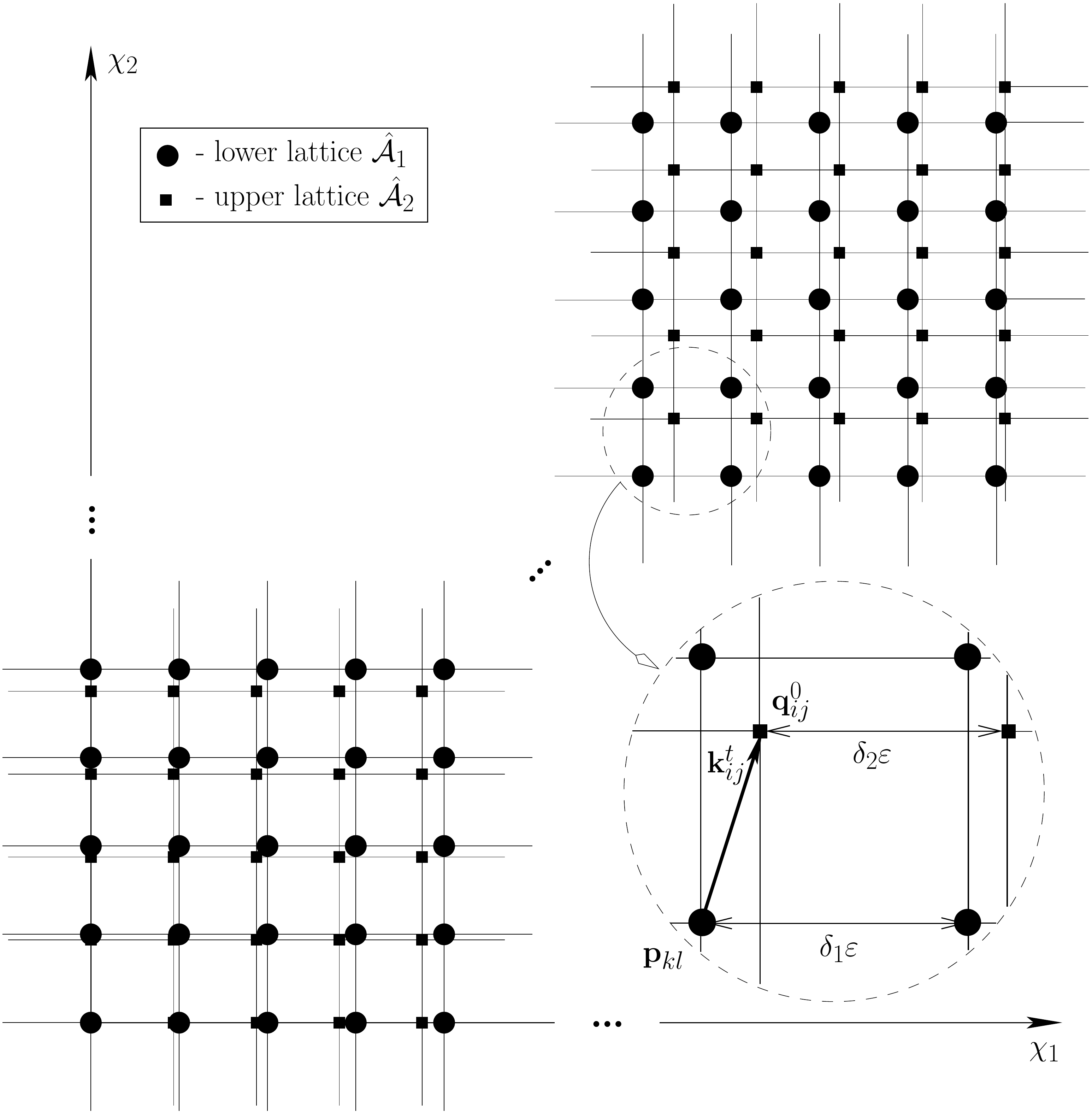}
  \caption{The reference configuration of the system of two square lattices ${\hat{\mathcal A}}_1$ and ${\hat{\mathcal A}}_2$ in nondimensional coordinates. The rigid lattice occupies an entire plane, while the deformable lattice is defined over the domain $[0,1]^2\subset\mathbb R^2$. Only two patches of the lattice structure are shown: near and away from the origin. The lattice parameters $\delta_1$ of ${\hat{\mathcal A}}_1$ and $\delta_2$ of ${\hat{\mathcal A}}_2$ are slightly different.}
  \label{f11}
\end{figure}

\begin{figure}[htb]
\centering
    \includegraphics[width=.6\linewidth]
                    {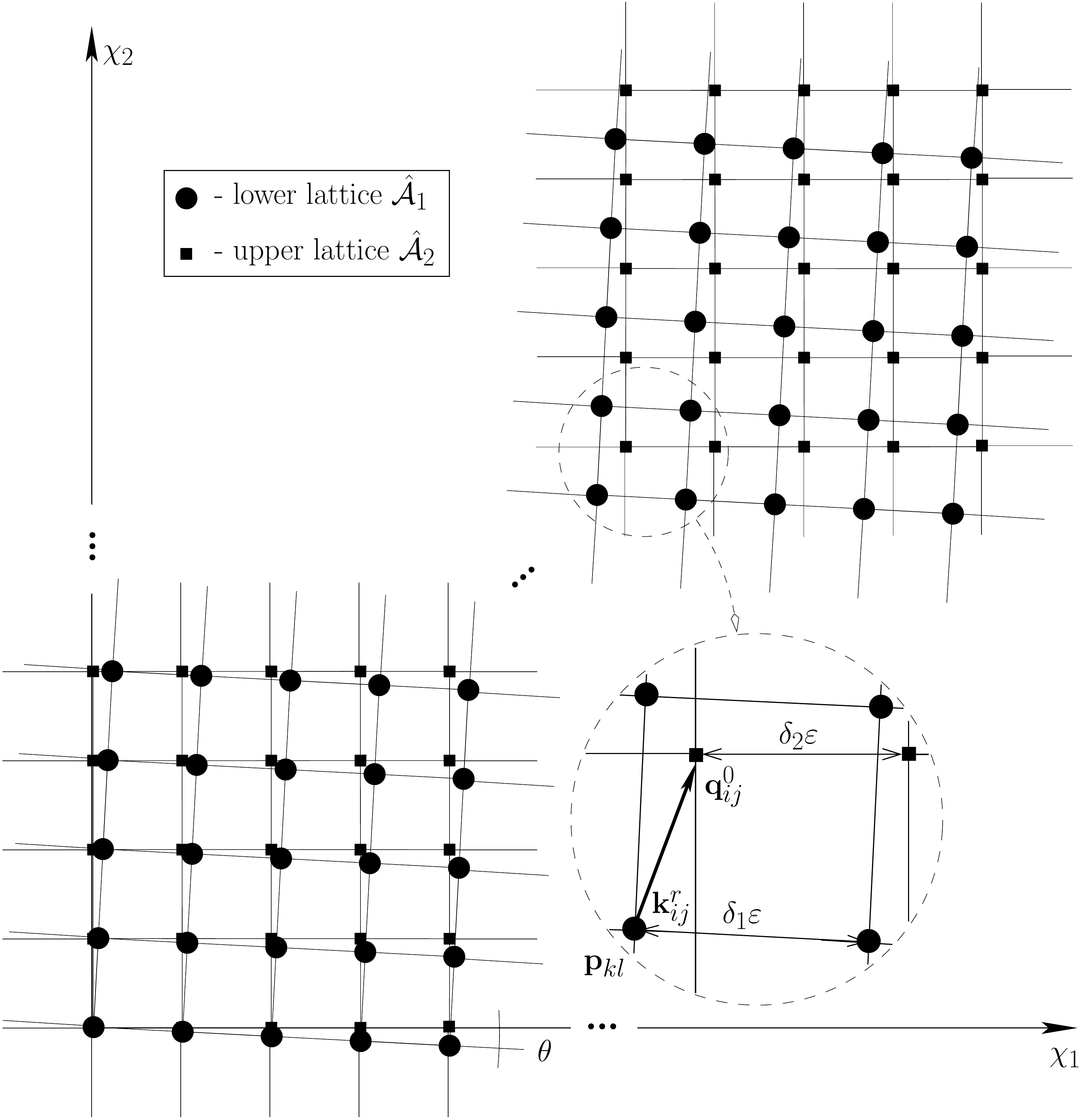}
    \caption{The reference configurations of the system of two square lattices ${\hat{\mathcal A}}_1$ and ${\hat{\mathcal A}}_2$ in nondimensional coordinates. The rigid lattice occupies an entire plane, while the deformable lattice is defined over the domain $[0,1]^2\subset\mathbb R^2$. Only two patches of the lattice structure are shown: near and away from the origin. The lattice orientations of ${\hat{\mathcal A}}_1$ and ${\hat{\mathcal A}}_2$ differ by a small angle $\theta$.}\label{ff35}
\end{figure}

To write down an expression for $d_{ij}^{mn}$, we let $\mathbf{k}_{ij}$ denote
the local horizontal mismatch between the atomic lattices $\hat{\mathcal{A}}_{1}$ and $\hat{\mathcal{A}}_{2}$ in the
{\it reference} configuration, as measured at the atom $i,j$ on $\hat{\mathcal{A}}_{2}$.
To determine $\mathbf{k}_{ij}$, we project the point ${\bf q}_{ij}^0$ onto the plane of the rigid lattice $\hat{\mathcal{A}}_{1}$. The projection falls inside one of the unit cells of $\hat{\mathcal{A}}_{1}$. Let $k,l$ be the indices of the lower left atom of this unit cell. We define ${\bf k}_{ij}$ to be a vector connecting the atom $k,l$ on $\hat{\mathcal{A}}_{1}$ to the projection of ${\bf q}_{ij}^0$ onto the plane of $\hat{\mathcal{A}}_{1}$ (see Figs. \ref{f11} and \ref{ff35}). 

To compute the local horizontal mismatch between the lattices in the {\it current} configuration, we recall that $\varepsilon\boldsymbol\xi\left(\eps\delta_2i,\eps\delta_2j\right)$ is the projection of the displacement vector for the atom $i,j$ onto the plane of $\hat{\mathcal{A}}_{2}$ in the reference configuration. It follows that the local horizontal mismatch in the current configuration is
\begin{equation}
{\bf k}_{ij}+\varepsilon\boldsymbol\xi\left(\eps\delta_2i,\eps\delta_2j\right).
\label{curroff}
\end{equation}

It is now clear that
\begin{equation}
  d_{ij}^{mn}
  =
  \left(\left\| \eps\delta_1(k-m)\,{\bf e}_1^1+\eps\delta_1(l-n)\,{\bf e}_2^1+\mathbf{k}_{ij}+\varepsilon\boldsymbol\xi\left(\eps\delta_2i,\eps\delta_2j\right)\right\|^{2}
  + \left(\varepsilon+\varepsilon\eta\left(\eps\delta_2i,\eps\delta_2j\right)\right)^{2}\right)^{\frac12},
  \label{ee15}
\end{equation}
where $\eps\delta_1(k-m)\,{\bf e}_1^1+\eps\delta_1(l-n)\,{\bf e}_2^1$ is the position of the atom $k,l$ on $\hat{\mathcal{A}}_{1}$ with respect to the atom $m,n$ on the same lattice. The expression $\varepsilon+\varepsilon\eta\left(\eps\delta_2i,\eps\delta_2j\right)$ gives the vertical distance between the lattices at the atom $i,j$ on $\hat{\mathcal{A}}_{2}$.

Recall that we postulated in \eqref{ee9} that the relative mismatch between the lattice parameters $\alpha\varepsilon$ is small. Assuming, in addition, that the angle $\theta$ of misorientation between two lattices is small, i.e.,
\begin{equation}
\Theta:=\frac{\theta}{\varepsilon}=O(1),
\label{angdef}
\end{equation}
we can linearize ${\bf k}_{ij}$ in $\varepsilon$ so that the relative contributions to ${\bf k}_{ij}$ from mismatch and misorientation can be computed separately and then added together. Consequently, we consider two possible choices for the reference configuration.

{\bf Case I - $\hat{\mathcal{A}}_{1}$ and $\hat{\mathcal{A}}_{2}$ have the same orientation, but different lattice parameters.} The corresponding reference configuration is shown in Fig.~\ref{f11}. Due to our assumption that an atom on $\hat{\mathcal{A}}_{2}$ at the origin lies directly above an atom on $\hat{\mathcal{A}}_{1}$, we have that
\begin{equation}
  {\bf k}_{ij}^t
  =
  \left(\text{mod}\,(i\varepsilon\delta_{2},\varepsilon\delta_{1}),\text{mod}\,(j\varepsilon\delta_{2},\varepsilon\delta_{1})\right)=(\delta_{2}-\delta_{1})\left(i\varepsilon,j\varepsilon\right)-\varepsilon\delta_1\left(i_1,j_1\right)
  =\varepsilon\alpha\left(i\varepsilon\delta_{2},j\varepsilon\delta_{2}\right)-\left(i_1\varepsilon\delta_1,j_1\varepsilon\delta_1\right),
  \label{ee4}
\end{equation}
where $\alpha$ is as defined in \eqref{ee9} and $i_1,j_1\in\mathbb N$. Because 
\begin{equation}
   \left(i\varepsilon\delta_{2},j\varepsilon\delta_{2}\right)=\boldsymbol\chi_{ij}, \label{ee5}
\end{equation}
the continuum approximation of ${\bf k}_{ij}^t$ is 
\begin{equation}
  {\bf k}_{ij}^t
 =
 \varepsilon\alpha\boldsymbol\chi_{ij} -\left(i_1\varepsilon\delta_1,j_1\varepsilon\delta_1\right).
  \label{eekt}
\end{equation}

{\bf Case II - $\hat{\mathcal{A}}_{1}$ and $\hat{\mathcal{A}}_{2}$ have different orientations, but the same lattice parameter.} The corresponding reference configuration is shown in Fig.~\ref{ff35}. From the figure and since $\delta_1=\delta_2$,
\begin{equation}
  {\bf k}_{ij}^r
  =
  \varepsilon\delta_1\left(i{\bf e}_1^2+j{\bf e}_2^2\right)-\varepsilon\delta_1\left(k{\bf e}_1^1+l{\bf e}_2^1\right)=\varepsilon\delta_1\left(i\left({\bf e}_1^2-{\bf e}_1^1\right)+j\left({\bf e}_2^2-{\bf e}_2^1\right)\right)+\varepsilon\delta_1\left(i_2{\bf e}_1^1+j_2{\bf e}_2^1\right),
  \label{ee4.1}
\end{equation}
where $i_2=i-k$ and $j_2=j-l$. The lattice $\hat{\mathcal{A}}_{1}$ is rotated with respect to $\hat{\mathcal{A}}_{2}$ by the angle $\theta$ with the corresponding rotation matrix
\[R(\theta)=
\left(
\begin{array}{ccc}
 \cos{\theta} & -\sin{\theta}   \\
 \sin{\theta} & \cos{\theta}
\end{array}
\right)
=\left(\begin{array}{ccc}
 \cos{\varepsilon\Theta} & -\sin{\varepsilon\Theta}   \\
 \sin{\varepsilon\Theta} & \cos{\varepsilon\Theta}
\end{array}
\right)
\sim\left(\begin{array}{ccc}
 1 & -\varepsilon\Theta  \\
 \varepsilon\Theta & 1
\end{array}
\right)
\]
with respect to the basis $\left\{{\bf e}_1^2,{\bf e}_2^2\right\}$, where we took into account \eqref{angdef}. It follows that
\[{\bf e}_1^2-{\bf e}_1^1={\bf e}_1^2-R(\theta)\,{\bf e}_1^2\sim-\varepsilon\Theta{\bf e}_2^2\]
and
\[{\bf e}_2^2-{\bf e}_2^1={\bf e}_2^2-R(\theta)\,{\bf e}_2^2\sim\varepsilon\Theta{\bf e}_1^2.\]
Inserting these expressions into \eqref{ee4.1} and using \eqref{ee5}, we find that the continuum approximation for ${\bf k}_{ij}^r$ is
\begin{equation}
{\bf k}_{ij}^r\sim\varepsilon\Theta\boldsymbol\chi_{ij}^\perp+\varepsilon\delta_1\left(i_2{\bf e}_1^1+j_2{\bf e}_2^1\right),
\label{ekr}
\end{equation}
where $\boldsymbol\chi^\perp=\left(\chi_2,-\chi_1\right)$.

We conclude that a continuum approximation of ${\bf k}_{ij}$ is 
\begin{equation}
  {\bf k}_{ij}
 \sim
 {\bf k}_{ij}^t+{\bf k}_{ij}^r
 =\varepsilon\alpha\boldsymbol\chi_{ij}+\varepsilon\Theta\boldsymbol\chi_{ij}^\perp+\left(\hat i\varepsilon\delta_1,\hat j\varepsilon\delta_1\right),
  \label{eek}
\end{equation}
where $\hat i=i_2-i_1$ and $\hat j=j_2-j_1$.

Substituting the expression for ${\bf k}_{ij}$ into \eqref{ee15} we obtain
\begin{multline*}
d_{ij}^{mn}
  \sim
  \varepsilon\left(\left\|\delta_1\left((k-m+\hat i)\,{\bf e}_1^1+(l-n+\hat j)\,{\bf e}_2^1\right)+\alpha\boldsymbol\chi_{ij}+\Theta\boldsymbol\chi^\perp_{ij}+\boldsymbol\xi\left(\boldsymbol\chi_{ij}\right)\right\|^{2}
  + \left(1+\eta\left(\boldsymbol\chi_{ij}\right)\right)^{2}\right)^\frac12 \\
  \sim  \varepsilon\left(\left\|\delta_2\left((k-m+\hat i)\,{\bf e}_1^2+(l-n+\hat j)\,{\bf e}_2^2\right)+\alpha\boldsymbol\chi_{ij}+\Theta\boldsymbol\chi^\perp_{ij}+\boldsymbol\xi\left(\boldsymbol\chi_{ij}\right)\right\|^{2}
  + \left(1+\eta\left(\boldsymbol\chi_{ij}\right)\right)^{2}\right)^\frac12,
\end{multline*}
for finite $m,n\in\mathbb Z$ because the basis of $\hat{\mathcal{A}}_1$ is a small perturbation of the basis of $\hat{\mathcal{A}}_2$ due to smallness of $\theta$. Returning to \eqref{ee3}, we see that the function $G$  that gives a continuum description of the van der Waals energy arising from the local lattice mismatch is defined by
\begin{equation}
	G(\boldsymbol\chi,\boldsymbol\xi(\boldsymbol\chi),\eta(\boldsymbol\chi)):
        =\mathcal G\left(\alpha\boldsymbol\chi+\Theta\boldsymbol\chi^\perp+\boldsymbol\xi\left(\boldsymbol\chi\right),\eta\left(\boldsymbol\chi\right)\right),
          \label{gcont}
\end{equation}
where
\begin{equation}
\mathcal G({\bf p},t):=\sum_{m,n=-\infty}^{\infty}
        g\left(
          \sqrt{\left\|\delta_2m\,{\bf e}_1^2+\delta_2n\,{\bf e}_2^2+{\bf p}\right\|^{2}
  + \left(1+t\right)^{2}}
          \right)
          \label{calg}
\end{equation}
for every ${\bf p}\in\mathbb R^2$ and $t>-1$. With a slight abuse of notation, here we changed the indices $m\to k-m+\hat i$ and $n\to k-n+\hat j$, respectively. Note that the infinite sum in\eqref{calg} converges due to the rapid decay of $g$.

Finally, the nondimensional version of \eqref{eq:DiscreteE_w} takes the form
\begin{align}\label{eq:DiscreteE_wND}
  {\mathcal E}_w({\bf Q}) 
  =
  \eps\sum_{i,j=1}^{N_2} \sum_{m,n=-\infty}^{\infty}  g\left(\|\bq_{ij}-\bp_{mn} \|\right)\sim\frac{1}{\eps}\sum_{i,j=1}^{N_2}G(\boldsymbol\chi,\boldsymbol\xi,\eta)\eps^2\sim \frac{1}{\varepsilon}\int_{[0,1]^2} G\left(\boldsymbol\chi,\boldsymbol\xi,\eta\right)\,d\boldsymbol\chi=:\mathcal F^\eps_w[\boldsymbol\xi,\eta],
\end{align}  
thus establishing \eqref{vdwfin}. 

\subsection{Continuum Energy} \label{encon}

Putting together all the contributions, the system is described by the following continuum energy functional
\begin{equation}
	\mathcal F^\eps[\boldsymbol\xi,\eta] := 
        \frac{\eps}{2}\int_{{[0,1]}^2}f\left(D\left(\nabla\boldsymbol\xi\right)+\frac{\eps}{2}\nabla\eta\otimes\nabla\eta\right)\,d{\boldsymbol\chi}
       +
        \frac{\gamma_d\eps^3}{2}\int_{[0,1]^2} {\left|\nabla\nabla\eta\right|}^2\,d{\boldsymbol\chi}
        +
        \frac{1}{\varepsilon}\int_{[0,1]^2} G\left(\boldsymbol\chi,\boldsymbol\xi,\eta\right)\,d\boldsymbol\chi.
        \label{ee6}
\end{equation}
Here $D(A)=(A+A^T)/2$ is the symmetric part of $A$ for any $A\in M^{2\times2}$ and 
\[f\left(M\right)=\gamma_s\left(m_{11}^2+m_{22}^2\right)+2\gamma_tm_{12}^2,\]
for any $M=\left(
\begin{array}{cc}
m_{11}   &  m_{12} \\
m_{12}   &  m_{22}
\end{array}
\right)
\in M^{2\times2}_{sym}$.

Note that the elastic contribution to the energy \eqref{ee6} is like that of the F\"oppl--von K\'arm\'an theory. The corresponding variational problem is of Ginzburg-Landau type, where the minimizers are determined via a competition between the elastic energy and the potential energy, which has multiple wells associated with the low-energy commensurate regions. The system is forced to reside in these wells, with the sharp transition between the wells being smoothed out due to the penalty imposed by the elastic energy.  Consequently, we expect the minimizers of \eqref{ee6} to develop walls of characteristic width $\eps$.


Now let ${\bf b}_1(\nabla\boldsymbol\xi,\nabla\eta)$ and ${\bf b}_2(\nabla\boldsymbol\xi,\nabla\eta)$ denote the columns of the matrix $D\left(\nabla\boldsymbol\xi\right)+\frac{\eps}{2}\nabla\eta\otimes\nabla\eta$. The Euler-Lagrange equations for the functional $\mathcal F^\eps$ are
\begin{equation}
\hspace{-1mm}\left\{
\begin{aligned}
&	-\eps\,\mathrm{div}\left[K_1{\bf b}_1(\nabla\boldsymbol\xi,\nabla\eta)
\right]
        +
        \frac1\varepsilon G_{\xi_1}(\boldsymbol\chi,\boldsymbol\xi,\eta)
        =0, \\
&	-\eps\,\mathrm{div}\left(K_2{\bf b}_2(\nabla\boldsymbol\xi,\nabla\eta)
\right]
        +
        \frac1\varepsilon G_{\xi_2}(\boldsymbol\chi,\boldsymbol\xi,\eta)
        =0,\\
&	\eps^3\gamma_d\Delta^2\eta-\eps^2\,\mathrm{div}\left(K_1{\bf b}_1(\nabla\boldsymbol\xi,\nabla\eta)\cdot\nabla\eta,K_2{\bf b}_2(\nabla\boldsymbol\xi,\nabla\eta)\cdot\nabla\eta
\right)
        +
        \frac1\varepsilon G_{\eta}(\boldsymbol\chi,\boldsymbol\xi,\eta)
        =0.         
\end{aligned}
\right.
\label{e17}
\end{equation}
Here the first two equations describe the force balance in the plane of the deformable lattice, while the last equation is the vertical force balance. The anisotropy matrices $K_1$ and $K_2$ are given by
\begin{equation}
\label{eq:anis}
K_1=\left(
\begin{array}{cc}
\gamma_s  &  0 \\
0  &  \gamma_t
\end{array}
\right)\qquad\mathrm{and}\qquad K_2=\left(
\begin{array}{cc}
\gamma_t  &  0 \\
0  &  \gamma_s
\end{array}
\right).
\end{equation}

\section{Numerical Results} \label{s3}


In this section, we numerically solve the system of Euler-Lagrange equations \eqref{e17} subject to periodic boundary conditions to explore the behavior of minimizers of the continuum model. 

\subsection{Periodic Boundary Conditions} \label{subsper}
First, we identify the constraints on the dimensionless parameters of the problem guaranteeing that the rotated rigid lattice $\hat {\mathcal A}_1$ coincides with its periodic extension to the exterior of the unit square $[0,1]^2$. 
\begin{figure}[htb]
\centering
    \includegraphics[width=.4\linewidth]
                    {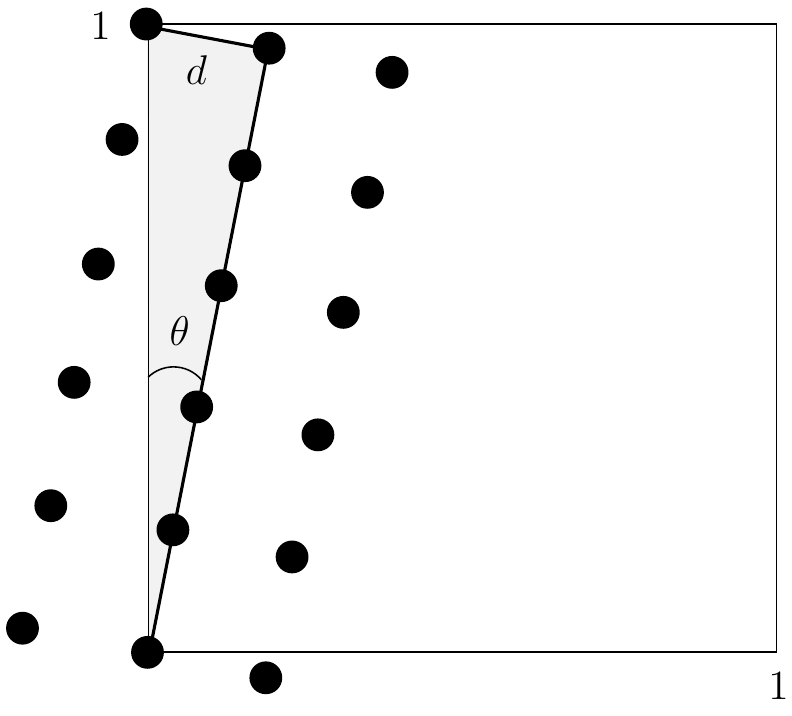}
    \caption{Rotated lattice with periodic boundary conditions on $[0,1]^2$. A sufficient condition for periodicity is that each corner of $[0,1]^2$ is occupied by an atom of the rotated lattice.}\label{fper}
\end{figure}
As suggested by Fig. \ref{fper}, a sufficient condition for periodicity is that each corner of $[0,1]^2$ is occupied by an atom of the rotated lattice. Using the shaded triangle in Fig. \ref{fper}, we conclude for the case shown that
\begin{equation}
n:=\frac{\cos{\theta}}{d}\in\mathbb N\qquad\mathrm{and}\qquad\sin{\theta}=d,
\label{eqper}
\end{equation}
where $n+1$ is the number of atomic rows that intersect each side of the square $[0,1]^2$. 

In fact, the second equation in \eqref{eqper} can be generalized via the following analogy. By rolling the square in Fig.~\ref{fper} into a tube with the axis parallel to, e.g., the horizontal side of the square, we obtain a tube with a lattice of atoms that can be placed on a helix. Following the helix around the tube for a single rotation, corresponds to advancing by one unit of length along the axis of the tube. We can say, for example, that the resulting helical structure has {\em chirality} $1$. Then, if this structure consisted of $k$ parallel equidistant helices with the axes that coincide with the axis of the tube, the chirality of the structure would have been equal $k$. Unwrapping the structure of any chirality back onto the unit square, would still produce a square lattice that can be periodically extended to the exterior of the square. It is not difficult to observe that the set of equations \eqref{eqper} can thus be written in a more general form
\begin{equation}
n:=\frac{\cos{\theta}}{d}\in\mathbb N\qquad\mathrm{and}\qquad\sin{\theta}=kd,\ k\in\mathbb N,
\label{eqper1}
\end{equation}
to incorporate lattices of any chirality. Note that, if the chirality $k$ is equal $0$, then the lattice vectors are parallel to the sides of the square.

Based on \eqref{eqper1}, we use the following procedure to generate the parameters for the rigid lattice $\hat {\mathcal A}_1$. Recall that we denoted the number of atomic rows for the deformable lattice $\hat{\mathcal A}_2$ in $[0,1]^2$ by $N_2$, so that 
\begin{equation}
\varepsilon\delta_2N_2=1.
\label{N2}
\end{equation}
After selecting two small integers $m\in\mathbb Z$ and $k\in\mathbb N$, we set 
\begin{equation}
N_1=N_2-m
\label{Nper}
\end{equation}
and find the angle $\theta$ from
\begin{equation}
\theta=\mathrm{atan}\,\left(\frac{k}{N_1}\right)
\label{thper}
\end{equation}
and the lattice constant for $\hat{\mathcal A}_1$ from
\begin{equation}
\delta_1=\frac{1}{\eps\sqrt{N_1^2+k^2}}.
\label{delper}
\end{equation}
Both of these expressions follow trivially from \eqref{eqper1} by setting $n=N_1$ and using that $d=\eps\delta_1$ for $\hat{\mathcal A}_1$. 

The parameters $\delta_1$, $N_1$, and $\theta$ as given by \eqref{Nper}, \eqref{thper}, and \eqref{delper} are used in discrete simulations below. We are now in a position to determine the parameters $\alpha$ and $\Theta$ for the corresponding continuum simulations. Indeed, from \eqref{angdef} we have that 
\begin{equation}
\Theta=\frac{\theta}{\eps}=\frac{1}{\eps}\mathrm{atan}\,\left(\frac{k}{N_1}\right)=\frac{1}{\eps}\mathrm{atan}\,\left(\frac{\eps k\delta_2}{\eps\delta_2(N_2-m)}\right)=\frac{1}{\eps}\mathrm{atan}\,\left(\frac{\eps k\delta_2}{1-\eps\delta_2m}\right)\sim \delta_2k,
\label{Thper}
\end{equation}
using \eqref{N2}-\eqref{thper} and that $\varepsilon\delta_2N_2=1$. Further, 
\begin{equation}
\frac{\delta_1-\delta_2}{\delta_2}=\frac{1}{\eps\delta_2\sqrt{N_1^2+k^2}}-1=\frac{1}{\eps\delta_2\sqrt{(N_2-m)^2+k^2}}-1=\frac{1}{\sqrt{(1-\eps\delta_2m)^2+\eps^2\delta_2^2k^2}}-1\sim\eps\delta_2m,
\label{ae34}
\end{equation}
by \eqref{N2}, \eqref{Nper}, and \eqref{delper}. It follows from \eqref{ae34} that the parameter $\alpha$, defined in \eqref{ee9}, is given by
\begin{equation}
\alpha\sim\delta_2m.
\label{alper}
\end{equation}
In what follows we refer to the parameter $m$, which measures the difference between the number of rows of atoms in the two lattices, as the {\em disparity}. Note that one can think of $m$ and $k$ as the components of a Burgers vector $\boldsymbol\beta:=(m,k)$ and interpret a wall in a system of two lattices as an edge and a screw dislocation when $\boldsymbol\beta:=(m,0)$ and $\boldsymbol\beta:=(0,k)$, respectively \cite{dai2016twisted}.

\subsection{Comparison between the discrete and continuum simulations}
For the numerical simulations below, we used COMSOL \cite{comsol} to solve the system of partial differential equations \eqref{e17} of the continuum model and LAMMPS \cite{PLIMPTON19951} to minimize the discrete energy \eqref{eq:entot}. For the continuum simulations we utilized the dissipation-dominated (gradient flow) dynamics to drive the energy of the system toward a (possibly local) minimum. The same task was accomplished for the discrete system of atoms by performing molecular dynamics simulations at a sufficiently low temperature. Both sets of simulations were conducted assuming periodic boundary conditions in the plane of the rigid lattice with period $1$ in $\chi_1$- and $\chi_2$-directions. For the initial conditions we assumed that the deformable lattice is parallel to and at a distance $\eps$ from the rigid lattice in nondimensional coordinates. We set $\delta_2=1$ and chose $m=0$, $k=3$, and $\eps=0.0238$ to generate a periodic system using the equations in the previous subsection. The resulting lattices have the same lattice constants but they are slightly rotated with respect to each other. 

\begin{figure}[htp]
\centering
    \qquad\qquad\qquad\qquad\qquad\includegraphics[width=.3\linewidth]
                    {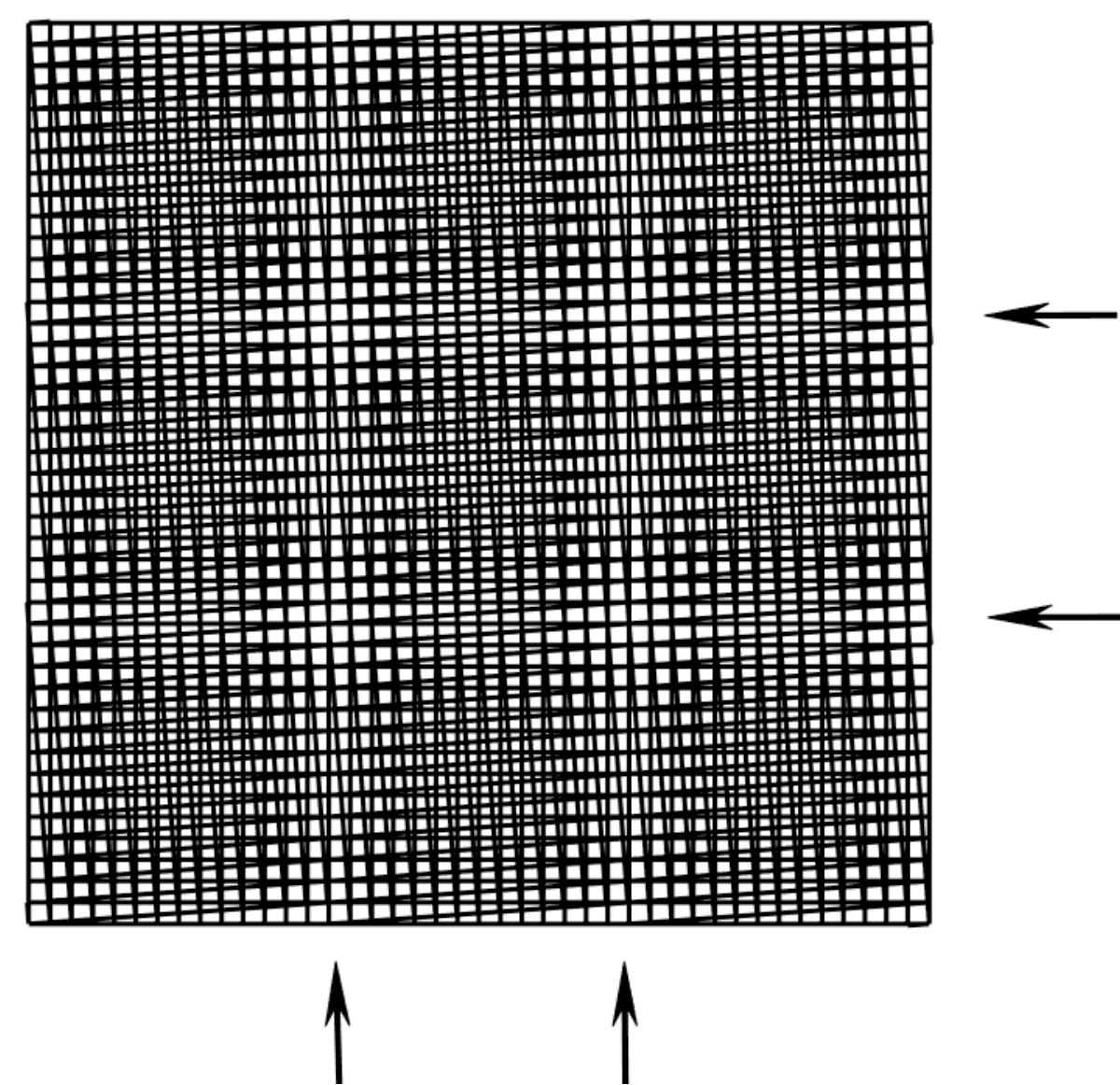}\hspace{1in}
  \caption{Moir\'e pattern for $m=0$, $k=3$, and $\eps=0.0238$. The bands along which the van der Waals energy density is the highest for the given initial spacing are indicated by arrows.}
  \label{fmoire}
\end{figure}

This periodic system has the moir\'e pattern depicted in Fig.~\ref{fmoire}, where the brighter regions indicated by arrows correspond to the least optimal registry from the point of view of the weak interaction. Indeed, this interaction prefers an atom of the deformable lattice to lie above the midpoint of a unit cell of the rigid lattice. One expects that, if the deformable lattice is allowed to relax, the darker regions would slightly rotate to increase the relative area of optimal registry. This should incur large elastic costs in the lighter regions of suboptimal registry, either by in-plane shear and/or by out-of-plane displacement.

\begin{figure}[H]
\centering
    \includegraphics[width=.4\linewidth]
                    {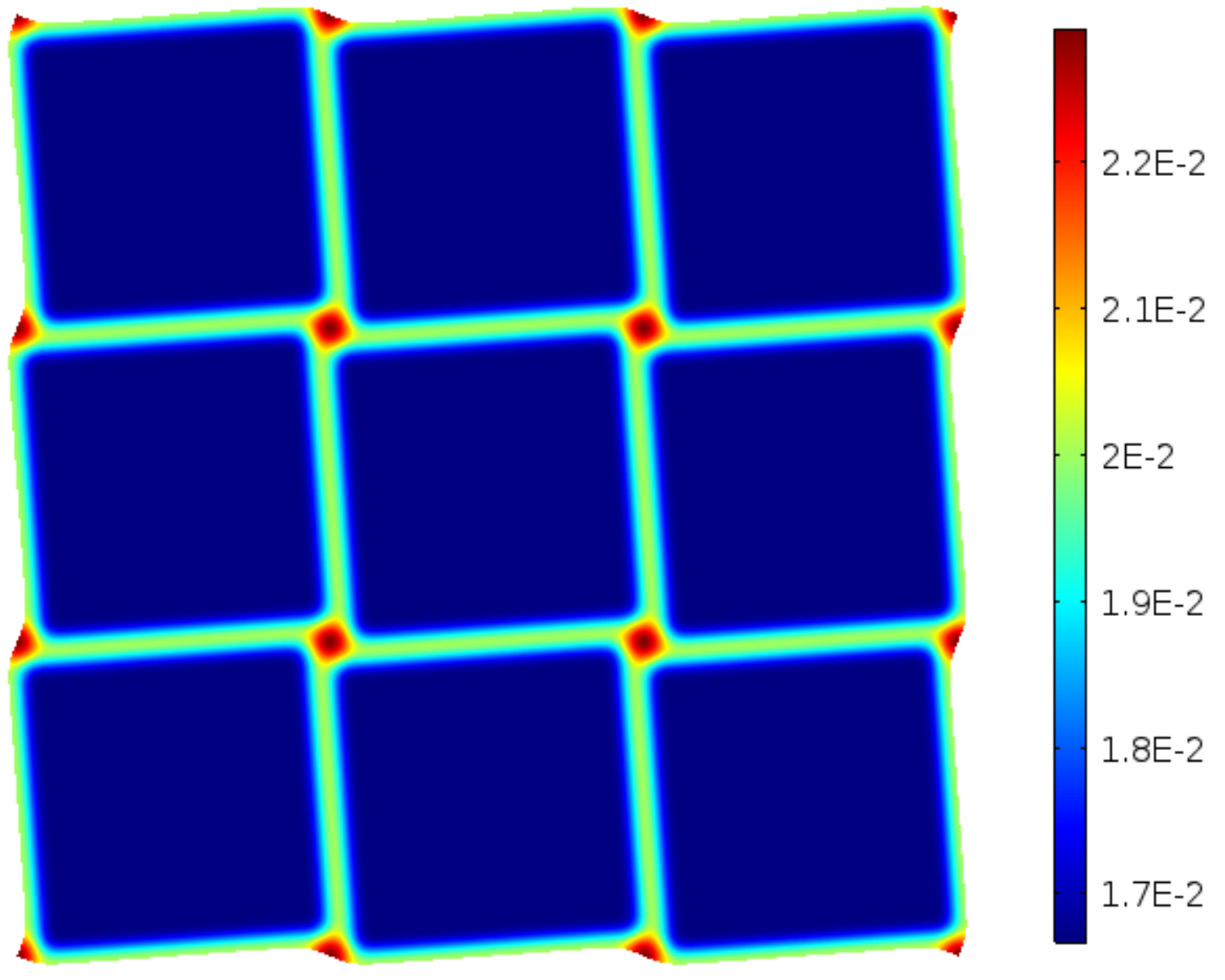}\qquad
                        \includegraphics[width=.43\linewidth]
                    {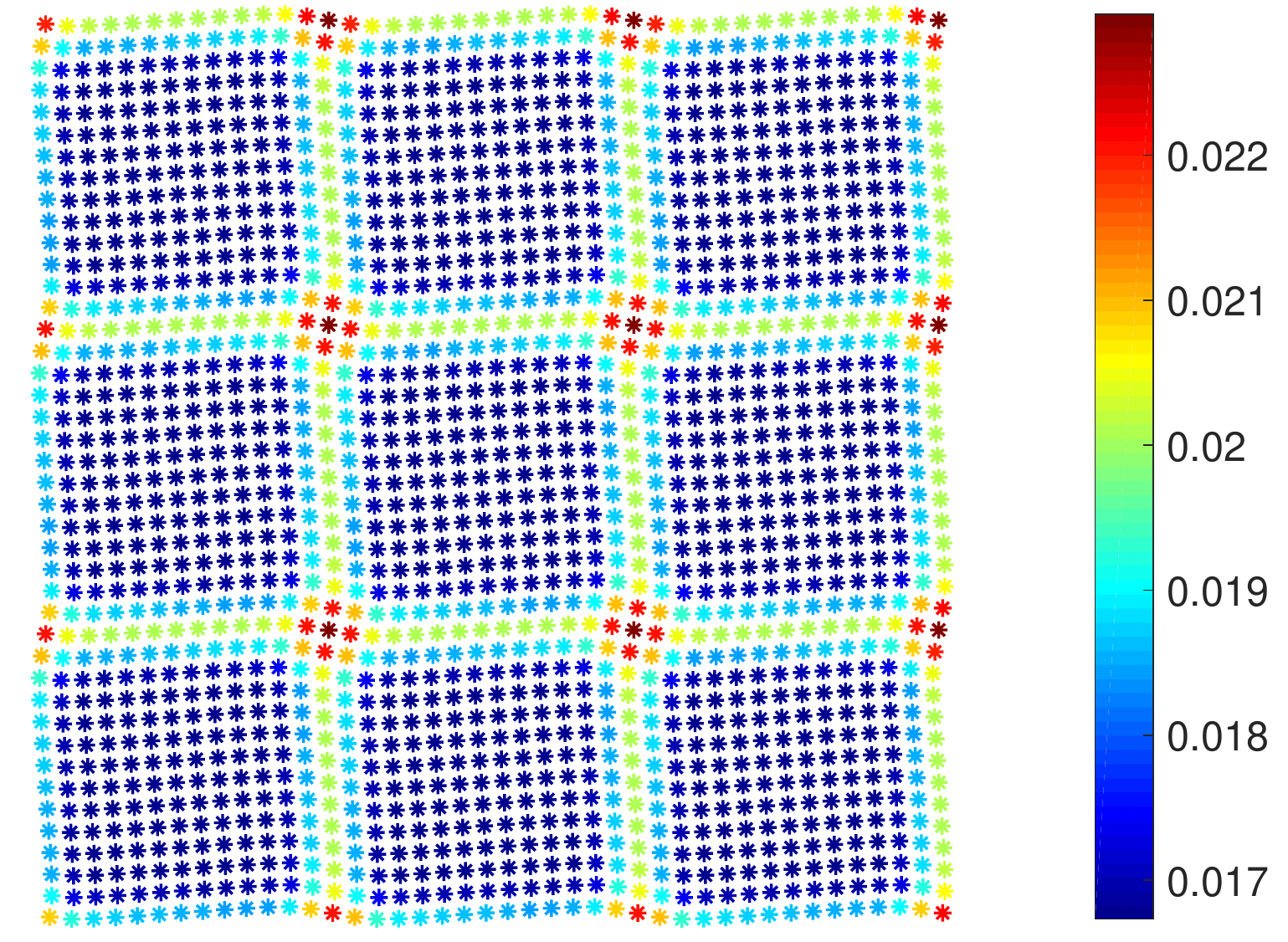}
  \caption{Continuum (left) and discrete (right) simulation results corresponding to the moir\'e pattern in Fig.~\ref{fmoire}.}
  \label{fcondis}
\end{figure}

We picked $\gamma_s=12$, $\gamma_t=16$, and $\gamma_d=.1$ for the elastic constants and determined the spring constants of the discrete model from \eqref{ee1.1} assuming that $\omega=1$. The results of the corresponding discrete and continuum simulations are shown in Fig.~\ref{fcondis}. 

The displacement field and the shear deformation observed in continuum simulations closely correspond to those determined for the discrete model. We observe that in both simulations the deformable lattice relaxes by generating six shear bands per period---three in the horizontal and three in the vertical directions, respectively. The shear bands coincide with the bright regions of the moir\'e pattern in Fig.~\ref{fmoire}, where the number of the bright regions is determined by the chirality $k$. The square-shaped regions of the deformable lattice between the shear bands rotate so that their orientations coincide with that of the rigid lattice. We refer to the shear bands as {\em domain walls} separating the regions of optimal lattice registry. 

Further inspection of Fig.~\ref{fcondis} shows that the domain walls also exhibit significant out-of plane displacements that reach their maximum values at ``hot spots"---the points of intersection between the walls. The minimum value of the displacement in Fig.~\ref{fcondis} corresponds to the equilibrium distance between an atom of the deformable lattice and the rigid lattice when this atom is positioned directly above the middle of a unit cell of the rigid lattice. In turn, the maximum value of the displacement corresponds to an atom at the equilibrium distance from the rigid lattice while positioned directly above an atom of that lattice. The displacements are maximized at hot spots, where the registry between the lattices is the least favorable from the point of view of weak interactions, while they are minimized in the square-shaped regions of the optimal registry. 

 \subsection{Pattern formation in continuum simulations}

In this subsection, we study pattern formation in the deformable lattice by solving the governing system \eqref{e17} for several combinations of geometric parameters of the model. In the remainder of this section each simulation is described by two figures. The first figure represents the deformed configuration with color scale indicating the local distance from the deformable to the rigid lattice. The second figure shows the deformation of a uniform grid drawn on the deformable lattice in the reference state, projected onto the $\chi_1\chi_2$-plane. Note that all lengths scale with $\eps$ and, in particular, the equilibrium distances depend on $\eps$.

\begin{figure}[htp]
\centering
    \includegraphics[width=.5\linewidth]
                    {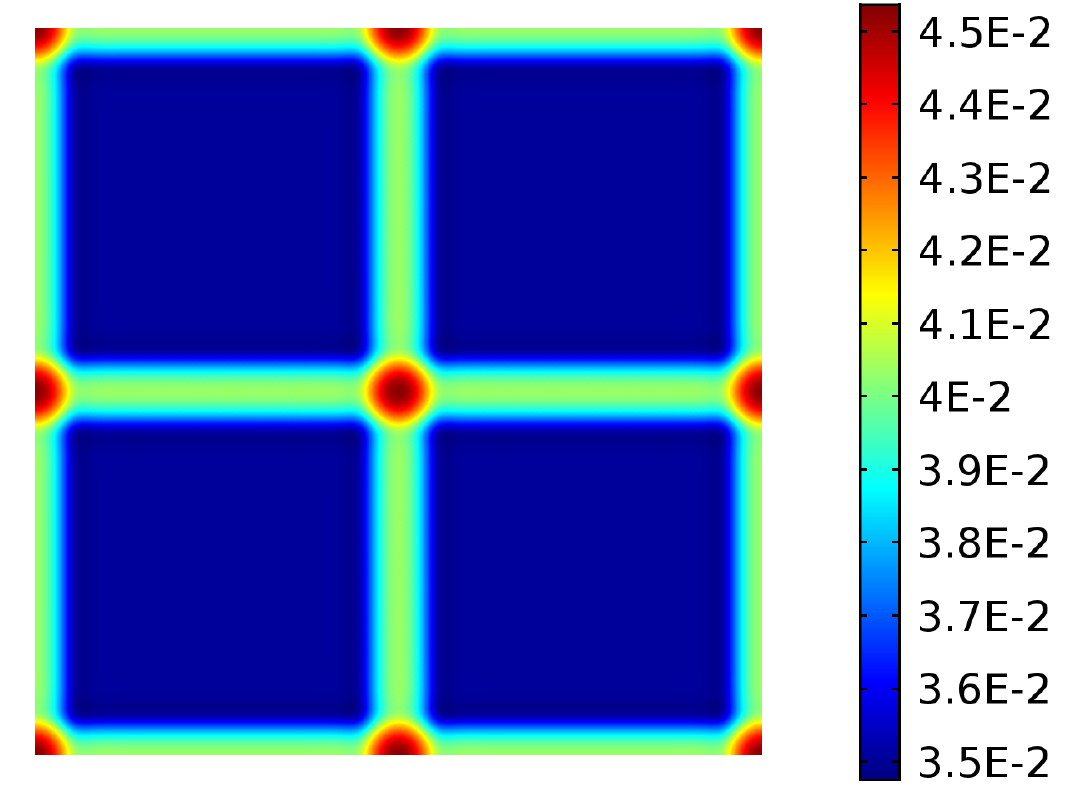}\hspace{.2in}
    \includegraphics[width=.36\linewidth]
                    {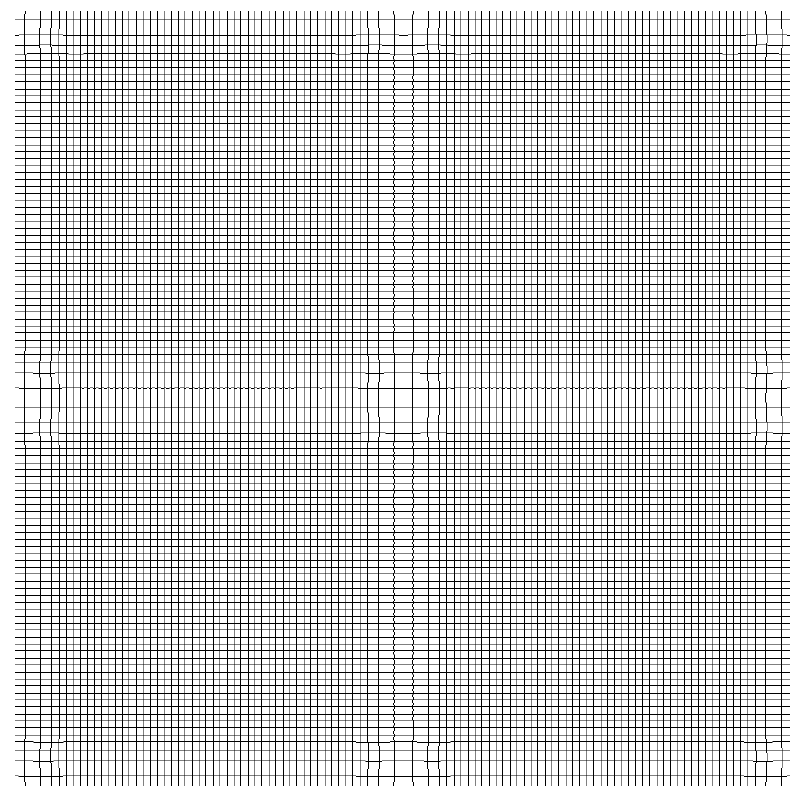}\hspace{.2in}
  \caption{Continuum simulations for $\eps=0.05$, $\gamma_s=1$, $\gamma_t=1$, $\gamma_d=0.1$, $m=2$, and $k=0$. Deformed configuration with the color indicating the local distance from the deformable to the rigid lattice (left) and the projection of the deformed uniform grid onto $\chi_1\chi_2$-plane (right).}
  \label{fc2}
\end{figure}

Suppose first that $\eps=0.05$, $\gamma_s=1$, $\gamma_t=1$, $\gamma_d=0.1$, $m=2$, and $k=0$. Since $k=0$, both the deformable and the rigid lattices have exactly the same orientation. At the same time, their lattice parameters are different so that the deformable lattice has exactly two extra vertical rows and two extra horizontal rows of atoms per period. The shape of the relaxed deformable lattice is shown in Fig.~\ref{fc2}. The system relaxes by pushing the extra rows of atoms slightly apart and farther away from the substrate to form the sets of two horizontal and two vertical walls within each period. Each wall accommodates exactly one extra row of atoms. Away from the walls, the deformable lattice slightly expands or contracts so that the atoms of both lattices are in optimal registry. 

\begin{figure}[htp]
\centering
    \includegraphics[width=.45\linewidth]
                    {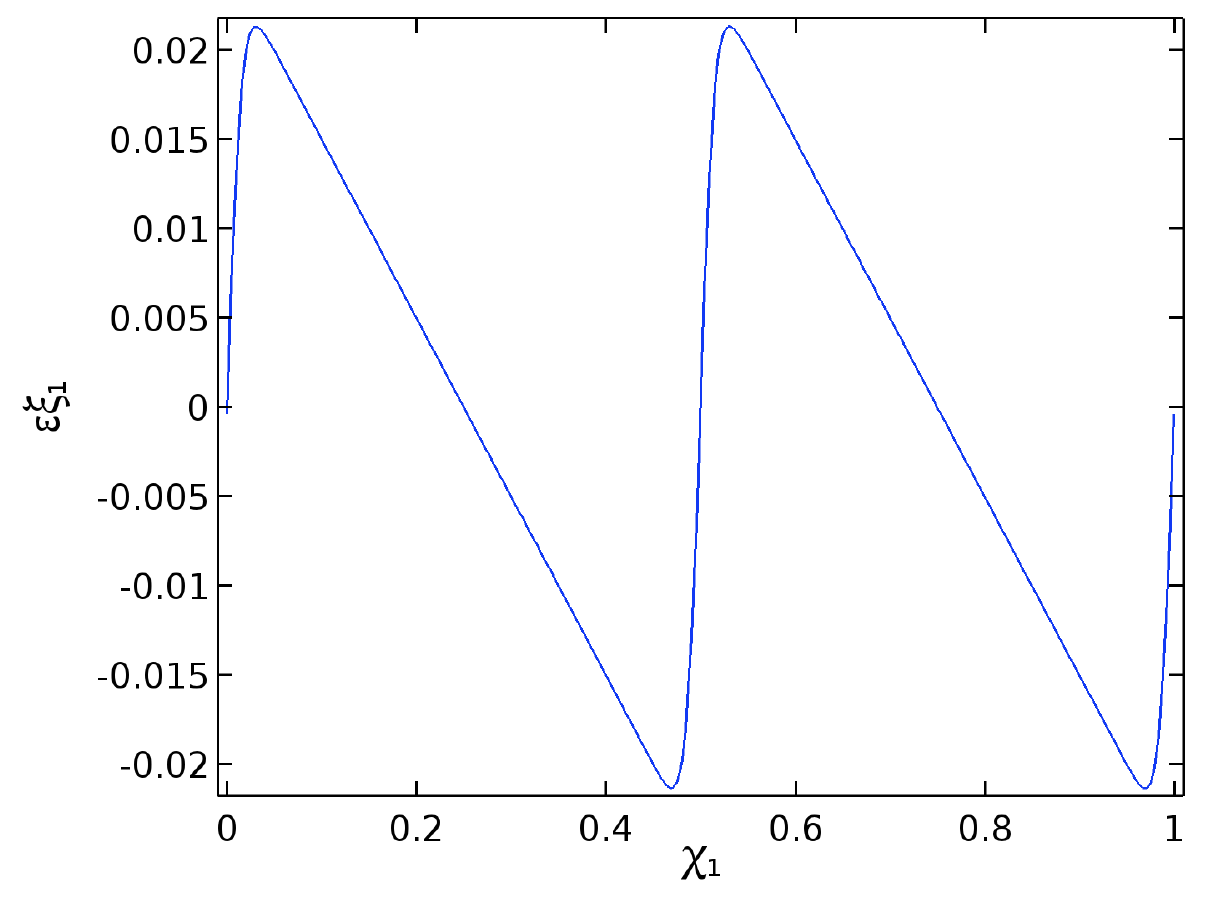}\hspace{.2in} 
                       \includegraphics[width=.45\linewidth]
                    {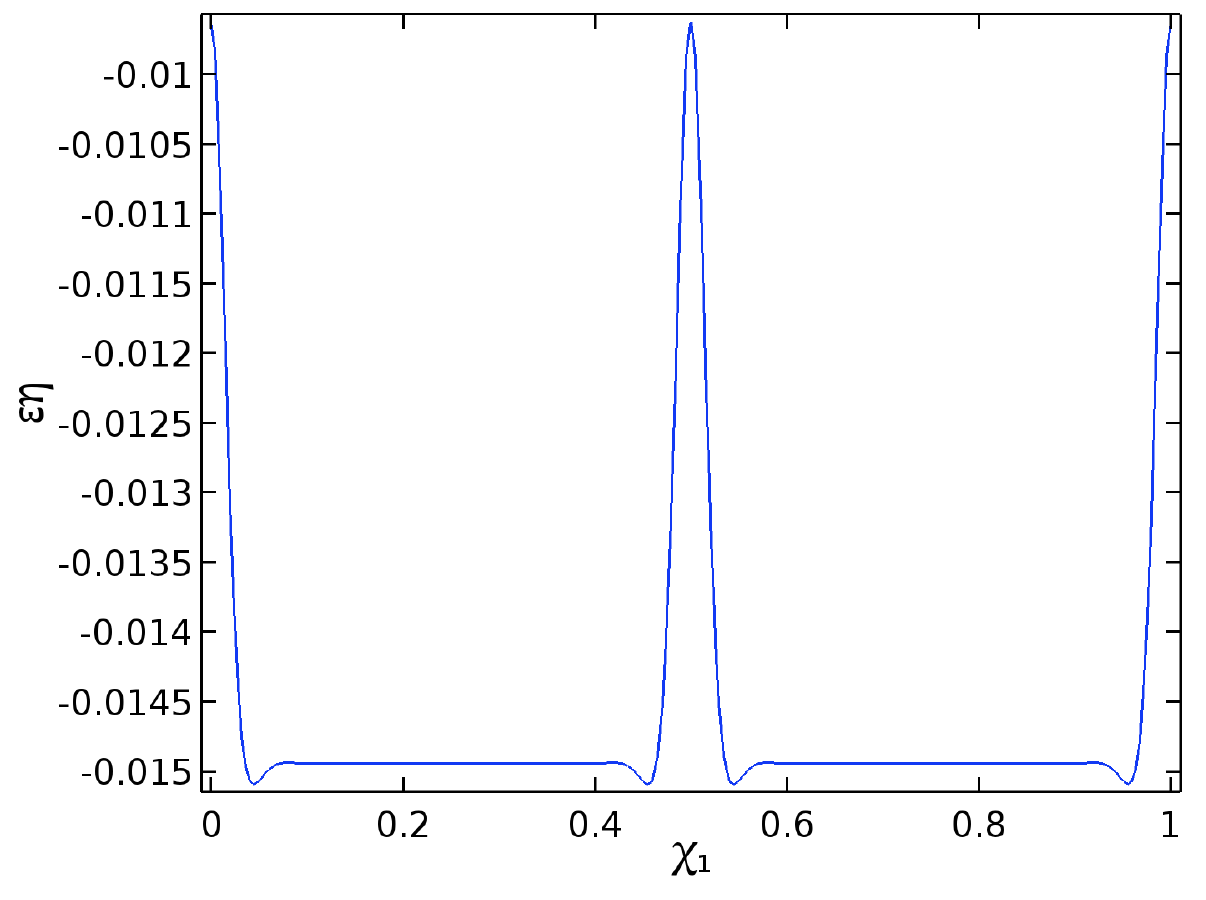}\hspace{1in}
  \caption{Continuum simulations for $\eps=0.05$, $\gamma_s=1$, $\gamma_t=1$, $\gamma_d=0.1$, $m=2$, and $k=0$. The dependence of $\eps\xi_1$ (left) and $\eps\eta$ (right) on $\chi_1$ when $\chi_2=0.25$.}
  \label{fc7}
\end{figure}

The corresponding nondimensional displacement components $\eps\xi_1$ and $\eps\eta$ are shown in Fig.~\ref{fc7} as functions of $\chi_1$ while holding $\chi_2=0.25$ fixed. These graphs are very similar to what is observed for an analogous one-dimensional system considered in \cite{2017arXiv170500072E}. In particular, the small dips next to the cross-section of a wall on the right inset in Fig.~\ref{fc7} result from the presence of the bending stiffness term and the linear stiffness of the van der Waals interaction near equilibrium. A macroscopic version of this effect is well known for elastic beams on a liquid \cite{PhysRevLett.107.044301}.

\begin{figure}[htp]
\centering
    \includegraphics[width=.5\linewidth]
                    {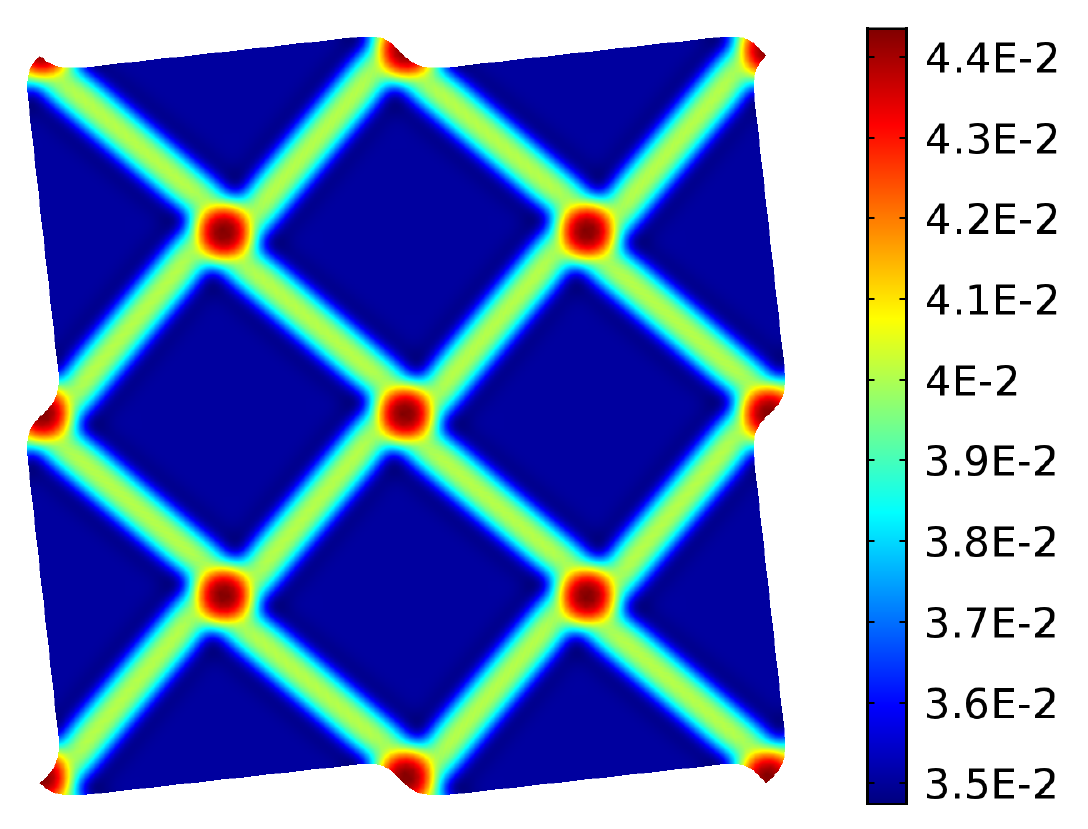}\hspace{.2in}
                     \includegraphics[width=.39\linewidth]
                    {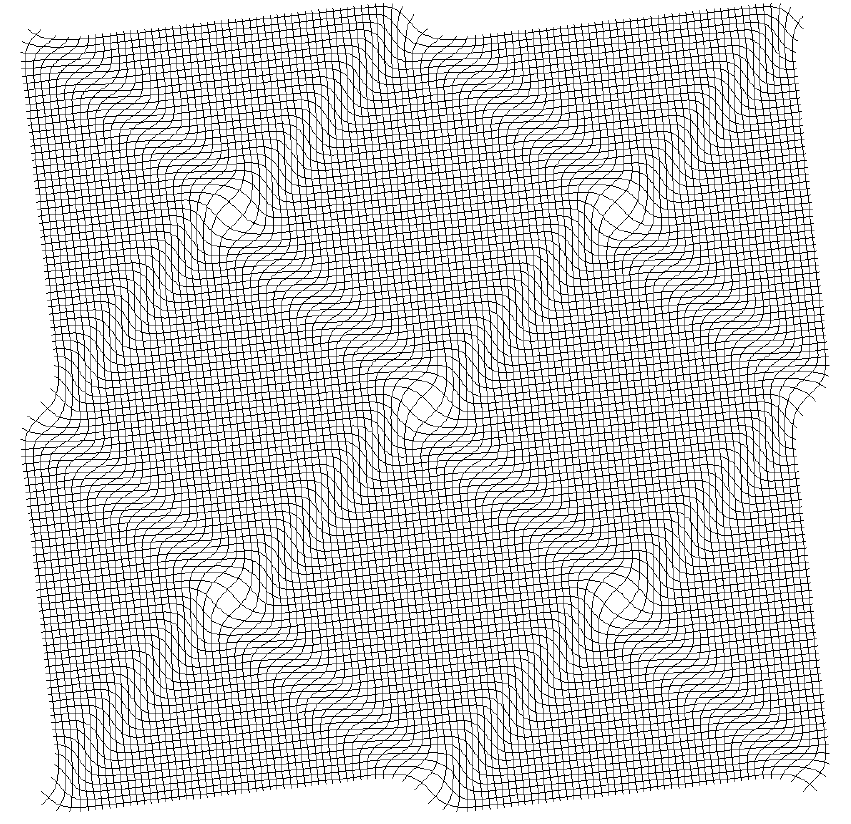}
  \caption{Continuum simulations for $\eps=0.05$, $\gamma_s=1$, $\gamma_t=1$, $\gamma_d=0.1$, $m=2$, and $k=2$. Deformed configuration with the color indicating the local distance from the deformable to the rigid lattice (left) and the projection of the deformed uniform grid onto the $\chi_1\chi_2$-plane (right).}
  \label{fc3}
\end{figure}

\begin{figure}[htp]
\centering
    \includegraphics[width=.5\linewidth]
                    {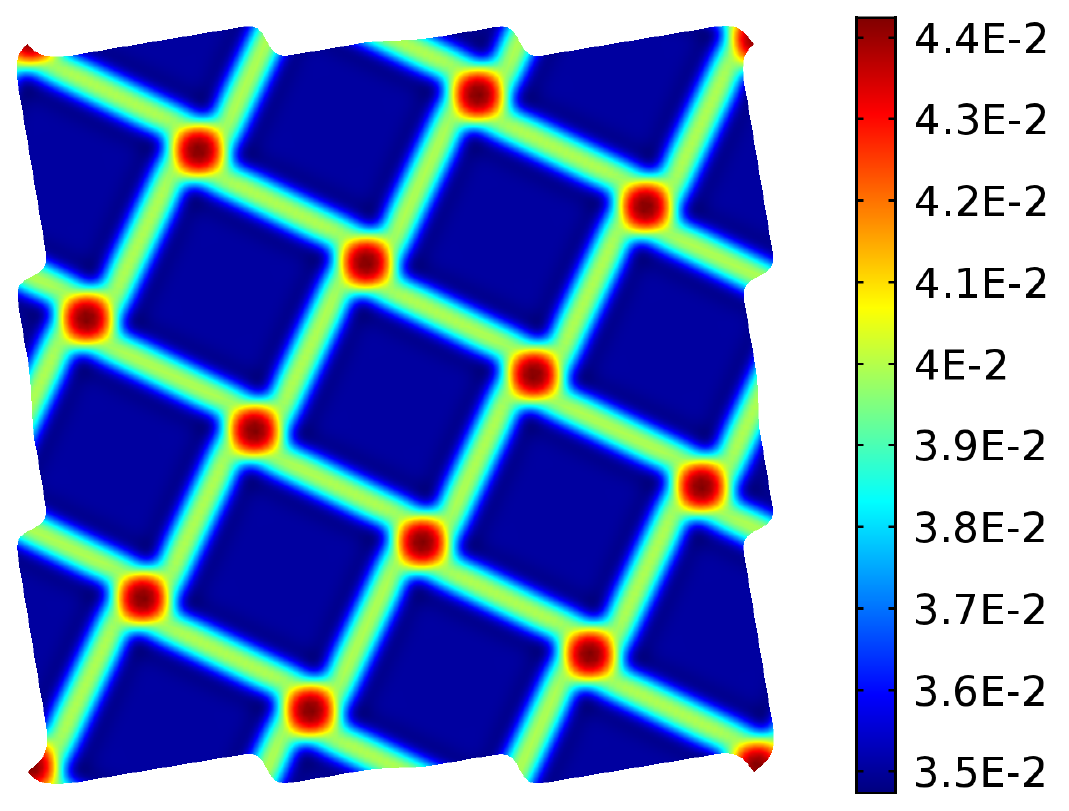}\hspace{.2in}
    \includegraphics[width=.39\linewidth]
                    {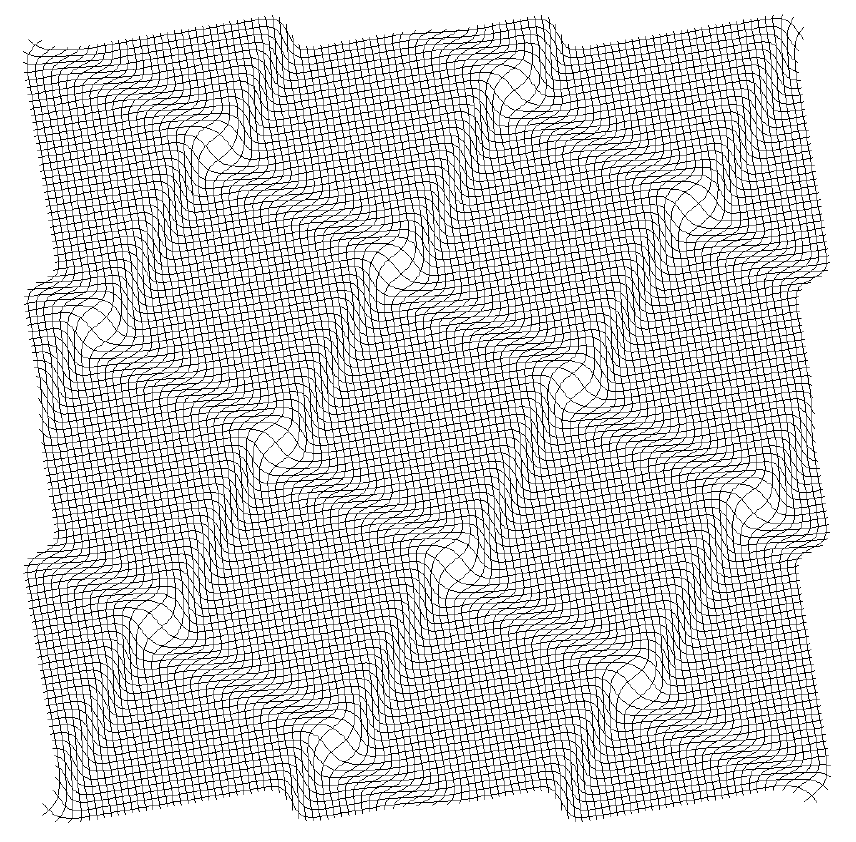}\hspace{.2in}
  \caption{Continuum simulations for $\eps=0.05$, $\gamma_s=1$, $\gamma_t=1$, $\gamma_d=0.1$, $m=2$, and $k=3$. Deformed configuration with the color indicating the local distance from the deformable to the rigid lattice (left) and the projection of the deformed uniform grid onto the $\chi_1\chi_2$-plane (right).}
  \label{fc4}
\end{figure}

\begin{figure}[htp]
\centering
    \includegraphics[width=.5\linewidth]
                    {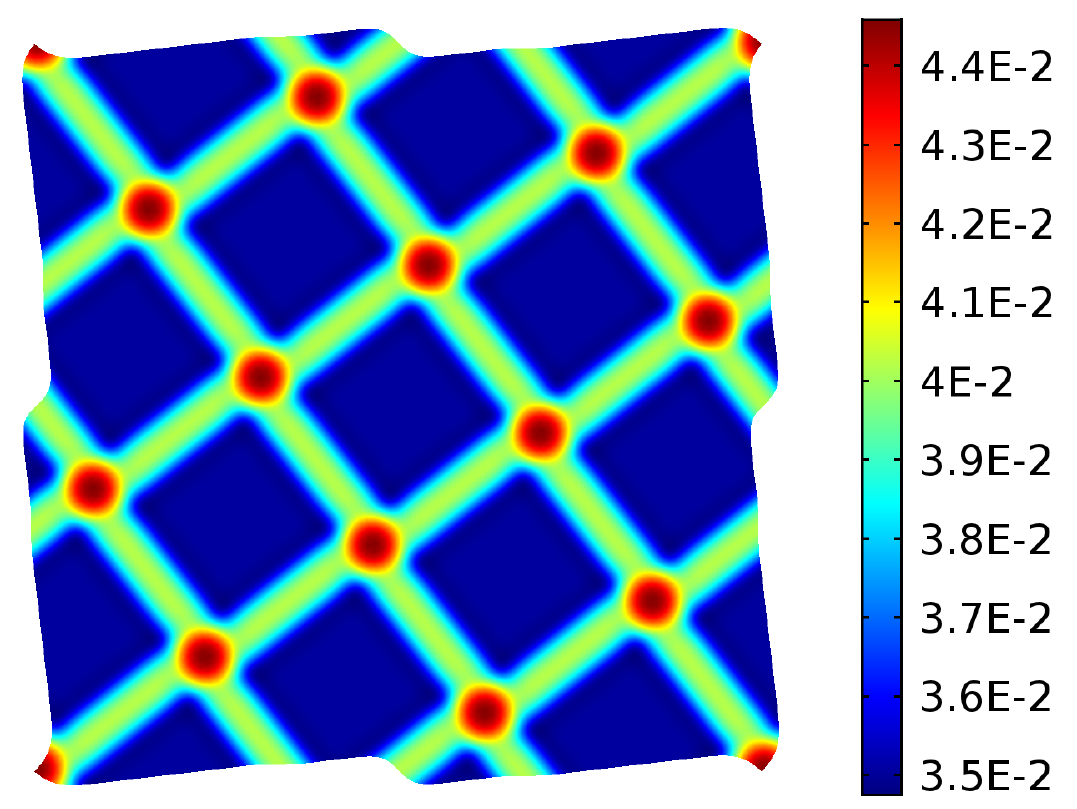}\hspace{.2in}
    \includegraphics[width=.39\linewidth]
                    {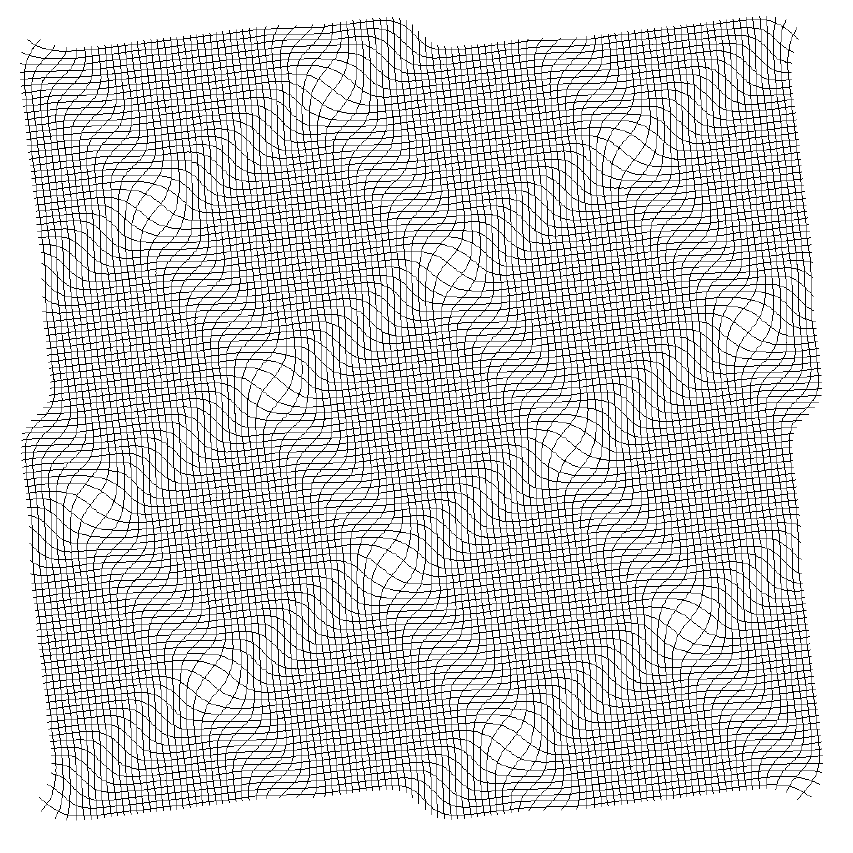}\hspace{.2in}
  \caption{Continuum simulations for $\eps=0.05$, $\gamma_s=1$, $\gamma_t=1$, $\gamma_d=0.1$, $m=3$, and $k=2$. Deformed configuration with the color indicating the local distance from the deformable to the rigid lattice (left) and the projection of the deformed uniform grid onto the $\chi_1\chi_2$-plane (right).}
  \label{fc5}
\end{figure}

We now explore the influence of the chirality $k$ and the disparity $m$ on the pattern of walls that form in the deformable lattice when both $k$ and $m$ are not equal zero. Examining Figs.~\ref{fc3}--\ref{fc5}, we observe that the number of walls that intersect each side of the square domain is given by the chirality $k$. There are two sets of parallel walls. For example, in Fig.~\ref{fc3}, the set of walls running from southeast to northwest corresponds to a stretch in the horizontal direction and a shear in the vertical direction. Conversely, the other set corresponds to a stretch in the vertical direction and a shear in the horizontal direction. Proceeding along any vertical line intersecting the square domain, one encounters exactly $m$ walls characterized by stretch in the vertical direction. Since the deformable lattice has $m$ extra rows of atoms compared to the rigid lattice, each wall serves to accommodate exactly one such row. A similar statement applies along for any horizontal line. Note also that out-of-plane deflection occurs along all walls.
\begin{figure}[htp]
\centering
    \includegraphics[width=.5\linewidth]
                    {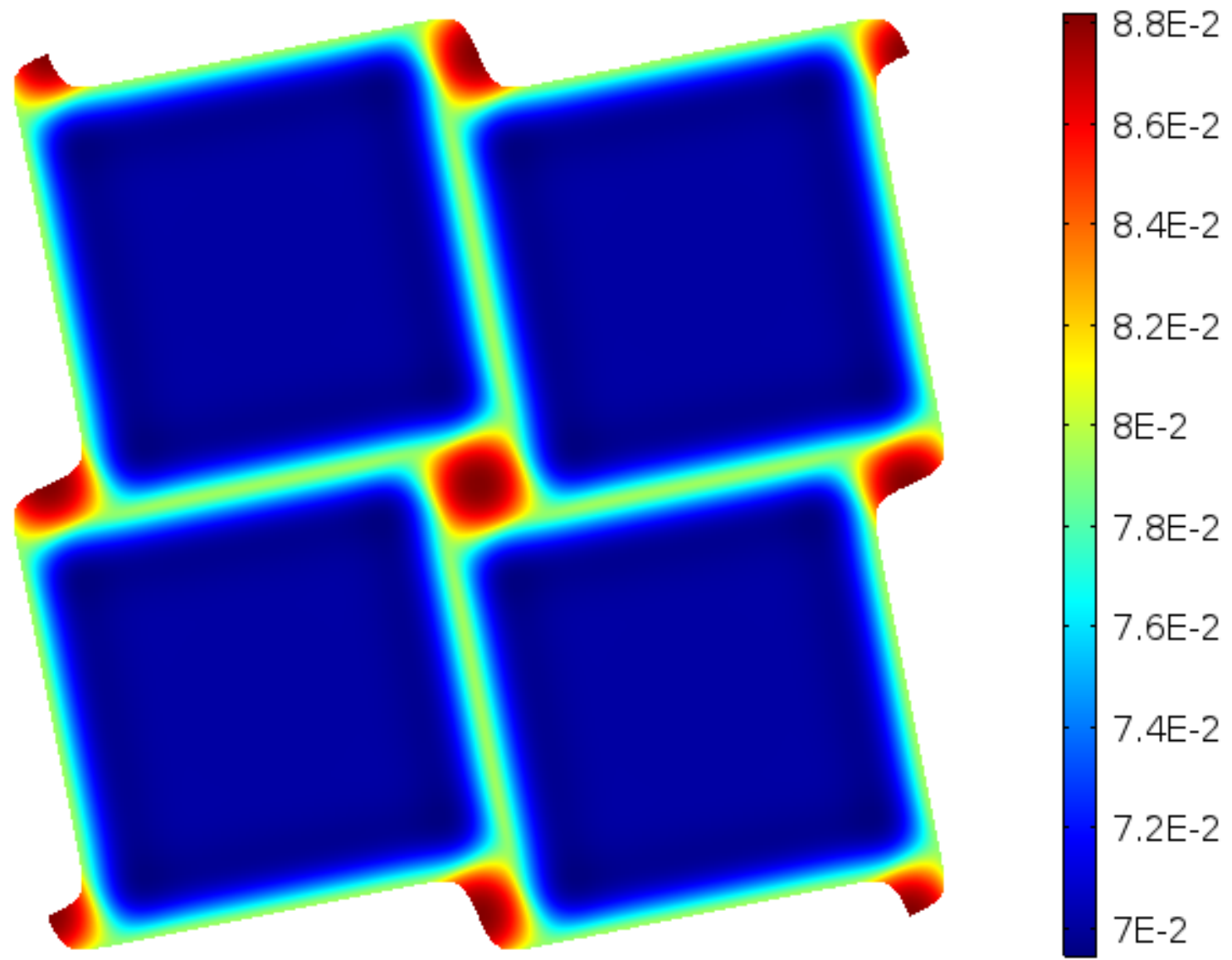}\hspace{.2in}
    \includegraphics[width=.39\linewidth]
                    {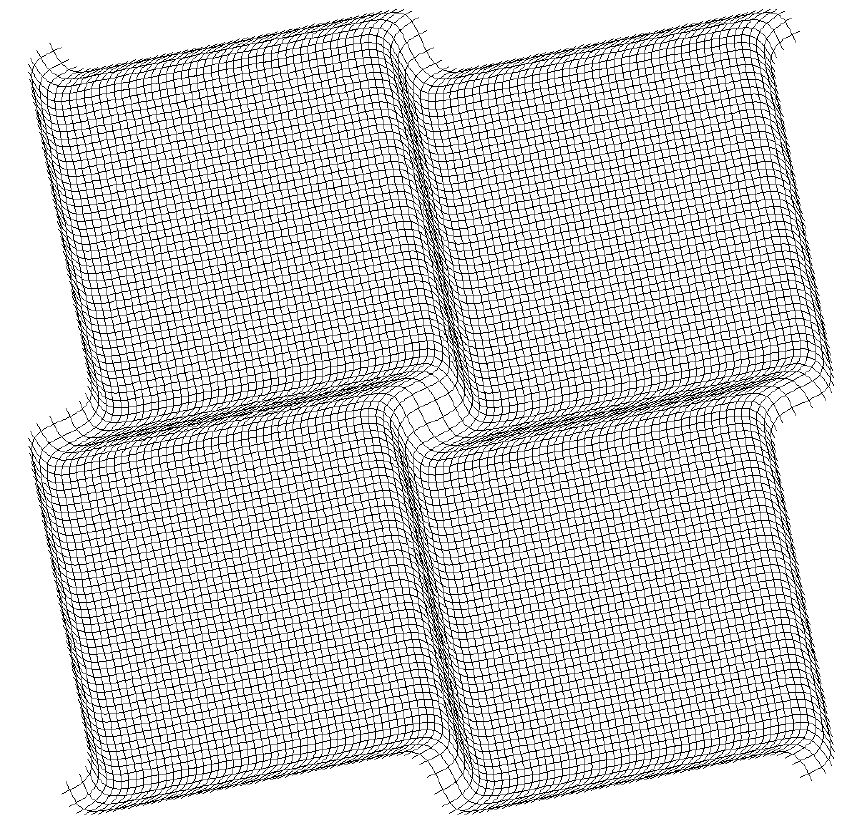}\hspace{.2in}
  \caption{Continuum simulations for $\eps=0.1$, $\gamma_s=1$, $\gamma_t=1$, $\gamma_d=0.1$, $m=0$, and $k=2$. Deformed configuration with the color indicating the local distance from the deformable to the rigid lattice (left) and the projection of the deformed uniform grid onto the $\chi_1\chi_2$-plane (right).}
  \label{fc6}
\end{figure}

Finally, increasing $\eps$ results in a larger angle of rotation between the rigid and deformable lattices that requires larger shear deformation to accommodate the weak interactions (see Fig.~\ref{fc6}).

\section{Conclusions}\label{sconc}

This work generalizes to two dimensions our previous results for weakly interacting Frenkel-Kontorova chains \cite{2017arXiv170500072E}. We have applied an upscaling procedure to develop a mesoscopic continuum model of a deformable two-dimensional lattice of atoms interacting with a rigid substrate.  We began with a system of atoms connected by harmonic extensional, torsional, and dihedral springs so that in equilibrium this system forms a square lattice. This lattice is assumed to weakly interact with another rigid square lattice via a van der Waals potential.  The upscaling procedure yields a continuum energy with terms describing the elastic energy of the deformable lattice and a term for the interaction energy between the deformable and rigid lattices.  Although a continuum description, the weak interaction energy retains discrete information about mismatch between the lattices. 

The numerical simulations were performed assuming periodic boundary conditions in the plane of the deformable lattice in the reference configuration. Simulations for a typical combination of geometric and material parameters demonstrate that the predictions of the mesoscopic model are in close correspondence with the configuration obtained via discrete molecular dynamics approach. Similar to \cite{dai2016structure}, we found that the deformable lattice develops a network of walls characterized by large shearing, stretching, and bending deformation that accommodate the misalignment and/or mismatch between the deformable and rigid lattices. We identified two integer-valued parameters describing the mismatch between the lattices. These parameters determine the geometry and the detail of deformation associated with the walls.  At the intersection of the walls, we find ``hot spots" characterized by large out-of-plane deformation \cite{van2014moire}.

Although the modeling presented here deals with square lattices, our approach is not limited to this choice of a system. Indeed, our procedure admits a straightforward generalization to any lattice type or atomic interactions. The periodic boundary conditions were imposed for simplicity and other boundary conditions can be considered. The model can also be extended in a standard way to include external body forces.

An interesting mathematical problem that arises from our upscaling procedure is to justify the number of terms that we retained in the asymptotic expansion of the elastic energy in $\eps$ in order to produce the continuum model. As discussed in Section \ref{elencon}, the terms we neglected influence the behavior of the solution inside the walls, where the deformation gradient is large. Our conjecture, confirmed by comparison with discrete simulations, is that the influence of the neglected terms on the network of walls, and by extension, on the solution in the commensurate regions is small. Indeed, our continuum model itself can be thought of as a truncation of an expansion of a continuum energy in terms of the same small parameter $\eps$. We expect that both the discrete and the continuum models approach in some appropriate sense the same ``limiting" model as $\eps$ tends to zero.  The exact framework and rigorous study of this convergence is a subject of future work. 

Continuum modeling that retains discrete registry effects is important for both solving computational problems more quickly and allowing theoretical insight into mesoscopic pattern formation of bilayer graphene and two-dimensional heterostructures. Further, continuum modeling may facilitate the study of the influence of atomic relaxation of slightly mismatched heterostructures on electronic properties of the system \cite{van2015relaxation}. 

\section{Acknowledgment}
This work was supported by the National Science Foundation grant DMS-1615952.

\section{Appendix}
\label{appe1}

In the Appendix, we expand in $\varepsilon$ all contributions to the discrete energy that led to the asymptotic expressions \eqref{stretch}--\eqref{dihe} in Section \ref{elencon}.  In what follows, we set ${\bf Q}=\left\{\bq_{ij}\right\}_{i,j=1}^{N_2}$, where $\bq_{ij}$ is given by \eqref{qij}.

\subsection{Extensional Springs}

We expand the first term in \eqref{eq:DiscreteE_s} in $\varepsilon$.  We first expand
$\bb^1_{ij}$.  We have
\begin{align}
  \bb^1_{ij}
  &=
  \bq_{i+1j}-\bq_{ij} \nonumber \\
  &=
  \left(
  \varepsilon\delta_{2}(i+1,j)
  +
  \varepsilon\boldsymbol{\xi}({\boldsymbol\chi}_{i+1j}),\,
  \varepsilon
  +
  \varepsilon\eta({\boldsymbol\chi}_{i+1j})  
  \right)
  -
  \left(
  \varepsilon\delta_{2}(i,j)
  +
  \varepsilon\boldsymbol{\xi}({\boldsymbol\chi}_{ij}),\,
  \varepsilon
  +
  \varepsilon\eta({\boldsymbol\chi}_{ij})  
  \right)  
  \nonumber \\
  &=
  \varepsilon\left(
  \delta_{2}(1,0)
  +
  \boldsymbol{\xi}_{,1}\varepsilon\delta_{2}
  +
  \frac{1}{2}\boldsymbol{\xi}_{,11}\varepsilon^{2}\delta_{2}^{2}
  +
  \frac{1}{6}\boldsymbol{\xi}_{,111}\varepsilon^{3}\delta_{2}^{3},\,
  \eta_{,1}\varepsilon\delta_{2}
  \right. \nonumber \\ & \qquad\qquad\qquad\qquad\qquad\qquad\qquad\qquad\left.+
  \frac{1}{2}\eta_{,11}\varepsilon^{2}\delta_{2}^{2}
  +
  \frac{1}{6}\eta_{,111}\varepsilon^{3}\delta_{2}^{3}
  \right)
  +
  \mathcal{O}(\varepsilon^{4}). 
  \label{ede1}
\end{align}
(Note that in \eqref{ede1} and in the expressions that follow, all
partial derivatives are evaluated at ${\boldsymbol\chi}_{ij}$.) 

We next expand
\begin{align}
  \|\bb^1_{ij}\|
  &=
  \delta_{2}\varepsilon
  +
  \xi_{1,1}\delta_{2}\varepsilon^{2}
  +
  \frac{1}{2}\left(\xi_{1,11}\delta_{2} + \eta_{,1}^{2} + \xi_{2,1}^{2} \right)
  \delta_{2}\varepsilon^{3} \nonumber \\
  &\phantom{mm}+
  \frac{1}{2}
  \left(
    \frac{1}{3}\xi_{1,111}\delta_{2}^{2}
    +
    \delta_{2}\eta_{,1}\eta_{,11}
    +
    \delta_{2}\xi_{2,1}\xi_{2,11}
    -
    \eta_{,1}^{2}\xi_{1,1}
    -
    \xi_{1,1}\xi_{2,1}^{2}
  \right)
  \delta_{2}\varepsilon^{4}
  +
  \mathcal{O}(\varepsilon^{5}).
  \label{ede22}
\end{align}

Lastly, we have
\begin{align}
  \frac{k_s}{2}\left( \frac{\|\bb^1_{ij}\|-\varepsilon\delta_{2}}{\varepsilon\delta_{2}}\right)^2
  &=
  \xi_{1,1}^{2}\varepsilon^{2}
  +
  \left(
    \delta_{2}\xi_{1,1}\xi_{1,11}
    +
    \eta_{,1}^{2}\xi_{1,1}
    +
    \xi_{1,1}\xi_{2,1}^{2}
  \right)
  \varepsilon^{3}
 +
  \mathcal{O}(\varepsilon^{4}).
  \label{ede23}
\end{align}

A similar computation for the second term in \eqref{eq:DiscreteE_s}
yields
\begin{align}
  \frac{k_s}{2}\left( \frac{\|\bb^2_{ij}\|-\varepsilon\delta_{2}}{\varepsilon\delta_{2}}\right)^2
  &=
  \xi_{2,2}^{2}\varepsilon^{2}
  +
  \left(
    \delta_{2}\xi_{2,2}\xi_{2,22}
    +
    \eta_{,2}^{2}\xi_{2,2}
    +
    \xi_{2,2}\xi_{1,2}^{2}
  \right)
  \varepsilon^{3}
  +
  \mathcal{O}(\varepsilon^{4}).
  \label{ede24}
\end{align}

We now combine \eqref{ede23} and \eqref{ede24}, which yields
\begin{multline}
   \mathcal E_s[\boldsymbol\xi,\eta]:=
  \sum_{i,j=1}^{N_2} \frac{k_s}{2\omega}
  \left[
    (\xi_{1,1}^{2}+\xi_{2,2}^{2})\varepsilon^{3}
    +
    \left(
    \delta_{2}(\xi_{1,1}\xi_{1,11} + \xi_{2,2}\xi_{2,22})
    \right.\right.\\\left.\left. +\,
    \eta_{,1}^{2}\xi_{1,1} + \eta_{,2}^{2}\xi_{2,2}
    +
    \xi_{1,1}\xi_{2,1}^{2} + \xi_{2,2}\xi_{1,2}^{2}
  \right)
  \varepsilon^{4}
  \right]
  +
  \mathcal{O}(\varepsilon^{5}).
  \label{ede28}
\end{multline}

\subsection{Torsional Springs}

We expand the first term in \eqref{e21.5} in $\varepsilon$.  To do
this, we use the expansions for $\bb^1_{ij}$ from \eqref{ede1}.  Also,
we expand $\bb^2_{ij}$ as
\begin{align}
  \bb^2_{ij}
  =
  \bq_{ij+1}-\bq_{ij} 
  &=
  \varepsilon\left(
  \delta_{2}(0,1)
  +
  \boldsymbol{\xi}_{,2}\varepsilon\delta_{2}
  +
  \frac{1}{2}\boldsymbol{\xi}_{,22}\varepsilon^{2}\delta_{2}^{2},\,
  \eta_{,2}\varepsilon\delta_{2}
  +
  \frac{1}{2}\eta_{,22}\varepsilon^{2}\delta_{2}^{2}
  \right)
  +\mathcal{O}(\varepsilon^{3}).
  \label{ede2}
\end{align}

Next we expand
\begin{equation}
  \bb^1_{ij}\cdot\bb^2_{ij}
  =
  \varepsilon^{3}\delta_{2}^{2}
  \left(
    \xi_{1,2}+\xi_{2,1}
    +
    \left(
      \frac{\delta_{2}}{2}(\xi_{1,22}+\xi_{2,11})
      -
      \nabla \xi_{1}\cdot \nabla \xi_{2}
      +
      \eta_{,1}\eta_{,2}
    \right)\varepsilon
    +  
    \mathcal{O}(\varepsilon^{2})
  \right), \label{ede17}
\end{equation}
so that
\begin{equation}
  \left(\bb^1_{ij}\cdot\bb^2_{ij}\right)^{2}
  =
  \varepsilon^{6}\delta_{2}^{4}
  \left(
    (\xi_{1,2}+\xi_{2,1})^{2}
     +
    2(\xi_{1,2}+\xi_{2,1})
    \left(
      \frac{\delta_{2}}{2}(\xi_{1,22}+\xi_{2,11})
      -
      \nabla \xi_{1}\cdot \nabla \xi_{2}
      +
      \eta_{,1}\eta_{,2}
    \right)\varepsilon
    +  
    \mathcal{O}(\varepsilon^{2})
  \right). \label{ede18}
\end{equation}
Also,
\begin{equation}
  \|\bb^1_{ij}\|^{2}
  =
  \varepsilon^{2}\delta_{2}^{2}
  \left(
    1
    +
    \mathcal{O}(\varepsilon^{2})
  \right),
  \qquad
  \|\bb^2_{ij}\|^{2}
  =
  \varepsilon^{2}\delta_{2}^{2}
  \left(
    1
    +
    \mathcal{O}(\varepsilon^{2})
  \right),
  \label{ede19}
\end{equation}
so that 
\begin{equation}
  (\|\bb^1_{ij}\|^{2}\|\bb^2_{ij}\|^{2})^{-1}
  =
  \left[
  \varepsilon^{4}\delta_{2}^{4}
  \left(
    1
    +
    \mathcal{O}(\varepsilon^{2})
  \right)
  \right]^{-1}
  =
  \varepsilon^{-4}\delta_{2}^{-4}
  \left(
    1
    +
    \mathcal{O}(\varepsilon^{2})
  \right).
  \label{ede20}
\end{equation}

Combining \eqref{ede18} and \eqref{ede20} yields
\begin{equation}
  \frac{\left(\bb^1_{ij}\cdot\bb^2_{ij}\right)^{2}}{\|\bb^1_{ij}\|^{2}\|\bb^2_{ij}\|^{2}}
  =
  \varepsilon^{2}
  \left(
    (\xi_{1,2}+\xi_{2,1})^{2}
   +
    2(\xi_{1,2}+\xi_{2,1})
    \left(
      \frac{\delta_{2}}{2}(\xi_{1,22}+\xi_{2,11})
      -
      \nabla \xi_{1}\cdot \nabla \xi_{2}
      +
      \eta_{,1}\eta_{,2}
    \right)\varepsilon
  \right)
  +  
  \mathcal{O}(\varepsilon^{4}).
  \label{ede21}
\end{equation}

Similar expansions yield
\begin{equation}
  \frac{\left(\bb^2_{ij}\cdot\bb^1_{i-1j}\right)^{2}}{\|\bb^2_{ij}\|^{2}\|\bb^1_{i-1j}\|^{2}}
  =
  \varepsilon^{2}
  \left(
    (\xi_{1,2}+\xi_{2,1})^{2}
    +
    2(\xi_{1,2}+\xi_{2,1})
    \left(
      \frac{\delta_{2}}{2}(\xi_{1,22}-\xi_{2,11})
      -
      \nabla \xi_{1}\cdot \nabla \xi_{2}
      +
      \eta_{,1}\eta_{,2}
    \right)\varepsilon
  \right) 
  +  
  \mathcal{O}(\varepsilon^{4}),
  \label{ede25.1}
\end{equation}
\begin{multline}
  \frac{\left(\bb^1_{i-1j}\cdot\bb^2_{ij-1}\right)^{2}}{\|\bb^1_{i-1j}\|^{2}\|\bb^2_{ij-1}\|^{2}}
  =
  \varepsilon^{2}
  \left(
    (\xi_{1,2}+\xi_{2,1})^{2}
    \right.\\\left.+
    2(\xi_{1,2}+\xi_{2,1})
    \left(
      \frac{\delta_{2}}{2}(-\xi_{1,22}-\xi_{2,11})
      -
      \nabla \xi_{1}\cdot \nabla \xi_{2}
      +
      \eta_{,1}\eta_{,2}
    \right)\varepsilon
  \right) 
  +  
  \mathcal{O}(\varepsilon^{4}),
  \label{ede25.2} 
\end{multline}
\begin{equation}
  \frac{\left(\bb^2_{ij-1}\cdot\bb^1_{ij}\right)^{2}}{\|\bb^2_{ij-1}\|^{2}\|\bb^1_{ij}\|^{2}}
  =
  \varepsilon^{2}
  \left(
    (\xi_{1,2}+\xi_{2,1})^{2}
    +
    2(\xi_{1,2}+\xi_{2,1})
    \left(
      \frac{\delta_{2}}{2}(-\xi_{1,22}+\xi_{2,11})
      -
      \nabla \xi_{1}\cdot \nabla \xi_{2}
      +
      \eta_{,1}\eta_{,2}
    \right)\varepsilon
  \right)
  +  
  \mathcal{O}(\varepsilon^{4}).
  \label{ede25.3}
\end{equation}

Combining \eqref{ede21} and \eqref{ede25.1}-\eqref{ede25.3}, we arrive at
\begin{equation}
  \mathcal E_t[\boldsymbol\xi,\eta]:=
  \sum_{i,j=1}^{N_2} \frac{k_t}{\omega}
  \left(
    2(\xi_{1,2}+\xi_{2,1})^{2}\varepsilon^{3}
   +
    4(\xi_{1,2}+\xi_{2,1})
      \left(
      -
      \nabla \xi_{1}\cdot \nabla \xi_{2}
      +
      \eta_{,1}\eta_{,2}
      \right)\varepsilon^{4}
  \right)
  +  
  \mathcal{O}(\varepsilon^{5}).    
  \label{ede26}
\end{equation}

\subsection{Dihedral Springs}

We expand the third term in \eqref{e_dih} in $\varepsilon$.  To do
this, we use the expansions for $\bb^1_{ij}$ and $\bb^2_{ij}$ from
\eqref{ede1} and \eqref{ede2}.  Also, we need the expansion for 
$\bb^1_{i-1j+1}$.  First, we write
\begin{align}
  \bb^1_{i-1j+1}
  &=
  \bq_{ij+1}-\bq_{i-1j+1} \nonumber \\
  &=
  (\bq_{ij+1}-\bq_{ij}) - (\bq_{i-1j+1}-\bq_{ij}). 
  \label{ede3}
\end{align}
The first term in \eqref{ede3} is $\bb^2_{ij}$.  For the second, we have 
\begin{align}
  \bq_{i-1j+1}-\bq_{ij} 
  &=
  \varepsilon\left(
  \delta_{2}(-1,1)
  +
  \boldsymbol{\xi}({\boldsymbol\chi}_{i-1j+1})-\boldsymbol{\xi}({\boldsymbol\chi}_{ij}),\,
  \eta({\boldsymbol\chi}_{i-1j+1})-\eta({\boldsymbol\chi}_{ij})    
  \right)  \nonumber \\
  &=
  \varepsilon\biggl(
  \delta_{2}(-1,1)
  +
  (-\boldsymbol{\xi}_{,1}+\boldsymbol{\xi}_{,2})\varepsilon\delta_{2}
  +
  \left(
    \frac{1}{2}\boldsymbol{\xi}_{,11}
    +
    \frac{1}{2}\boldsymbol{\xi}_{,22}
    -
    \boldsymbol{\xi}_{,12}
  \right)\varepsilon^{2}\delta_{2}^{2},\,
    \nonumber \\
  &\phantom{mmmmmm}
  (-\eta_{,1}+\eta_{,2})\varepsilon\delta_{2}
  +
  \left(
    \frac{1}{2}\eta_{,11}
    +
    \frac{1}{2}\eta_{,22}
    -
    \eta_{,12}
  \right)\varepsilon^{2}\delta_{2}^{2}
  \biggr)
  +\mathcal{O}(\varepsilon^{3}),
  \label{ede4}
\end{align}
Now subtracting \eqref{ede4} from the right-hand side of \eqref{ede2} yields
\begin{equation}
  \bb^1_{i-1j+1} 
  =
  \varepsilon\biggl(
  \delta_{2}(1,0)
  +
  \boldsymbol{\xi}_{,1}\varepsilon\delta_{2}
  +
  \left(
    \frac{1}{2}\boldsymbol{\xi}_{,11}
    +
    \boldsymbol{\xi}_{,12}
  \right)\varepsilon^{2}\delta_{2}^{2},\, \eta_{,1}\varepsilon\delta_{2}
  +
  \left(
    -\frac{1}{2}\eta_{,11}
    +
    \eta_{,12}
  \right)\varepsilon^{2}\delta_{2}^{2}
  \biggr)  
  +
  \mathcal{O}(\varepsilon^{3}).
  \label{ede5}
\end{equation}

Returning to \eqref{e_dih}, we now must expand
\begin{align}
  \bb^1_{ij}\times\bb^2_{ij}
  &=\varepsilon^{2}\delta_{2}^{2}
  \biggl(-\eta_{,1}\varepsilon
  +
  \left(\xi_{2,1}\eta_{,2}-\frac{1}{2}\eta_{,11}\delta_{2}-\xi_{2,2}\eta_{,1}\right)\varepsilon^{2},\,
  \nonumber \\
  &\phantom{mmmm}
  -\eta_{,2}\varepsilon
  +
  \left(\xi_{1,2}\eta_{,1}-\frac{1}{2}\eta_{,22}\delta_{2}-\xi_{1,1}\eta_{,2}\right)\varepsilon^{2},\,1
  +
  (\xi_{1,1}+\xi_{2,2})\varepsilon
  \nonumber \\
  &\phantom{mmmm}
    +
  \left(\frac{1}{2}(\xi_{1,11}+\xi_{2,22})\delta_{2}
   +\xi_{1,1}\xi_{2,2}-\xi_{1,2}\xi_{2,1}\right)\varepsilon^{2}\biggr)
  +
  \mathcal{O}(\varepsilon^{3}).
  \label{ede7}
\end{align}

Next we compute
\begin{align}
  (\bb^1_{ij}&\times\bb^2_{ij})\cdot \bb^1_{i-1j+1}
  = \nonumber \\
  &
  \varepsilon^{3}\delta_{2}^{3}
  \biggl[
    \left(-\eta_{,1}\varepsilon
    +
    \left(\xi_{2,1}\eta_{,2}-\frac{1}{2}\eta_{,11}\delta_{2}-\xi_{2,2}\eta_{,1}\right)\varepsilon^{2}
    +
    \mathcal{O}(\varepsilon^{3})\right)\times \nonumber \\
    &\phantom{mmmmmm}
    \left(\delta_{2}
    +
    \xi_{1,1}\varepsilon\delta_{2}
    +
    \frac{1}{2}\xi_{1,11}\varepsilon^{2}\delta_{2}^{2}
    +
    \xi_{1,12}\varepsilon^{2}\delta_{2}^{2}
    +
    \mathcal{O}(\varepsilon^{3})
    \right)
    \nonumber \\
    &\phantom{m}+
    \left(-\eta_{,2}\varepsilon
    +
    \left(\xi_{1,2}\eta_{,1}-\frac{1}{2}\eta_{,22}\delta_{2}-\xi_{1,1}\eta_{,2}\right)\varepsilon^{2}
    +
    \mathcal{O}(\varepsilon^{3})\right)\times \nonumber \\
    &\phantom{mmmmmm} 
    \left(
    \xi_{2,1}\varepsilon\delta_{2}
    +
    \frac{1}{2}\xi_{2,11}\varepsilon^{2}\delta_{2}^{2}
    +
    \xi_{2,12}\varepsilon^{2}\delta_{2}^{2}
    +
    \mathcal{O}(\varepsilon^{3})
    \right)
    \nonumber \\
    &\phantom{m}+
    \left(1
    +
    (\xi_{1,1}+\xi_{2,2})\varepsilon
    +
    \left(\frac{1}{2}(\xi_{1,11}+\xi_{2,22})\delta_{2}
    +\xi_{1,1}\xi_{2,2}-\xi_{1,2}\xi_{2,1}\right)\varepsilon^{2}
    +
    \mathcal{O}(\varepsilon^{3})\right)\times \nonumber \\
    &\phantom{mmmmmm} 
    \left(
    \eta_{,1}\varepsilon\delta_{2}
    -
    \frac{1}{2}\eta_{,11}\varepsilon^{2}\delta_{2}^{2}
    +
    \eta_{,12}\varepsilon^{2}\delta_{2}^{2}
    +
    \mathcal{O}(\varepsilon^{3}
    \right)
    \biggr]    
    \nonumber \\
    &=
    \varepsilon^{3}\delta_{2}^{3}
    \left[
      (\eta_{,12}-\eta_{,11})\varepsilon^{2}\delta_{2} + \mathcal{O}(\varepsilon^{3})
    \right].
  \label{ede8}
\end{align}
Hence
\begin{equation}
  \left[
    (\bb^1_{ij}\times\bb^2_{ij})\cdot \bb^1_{i-1j+1}
  \right]^{2}
  = 
  \varepsilon^{6}\delta_{2}^{6}
  \left[
    (\eta_{,12}-\eta_{,11})^{2}\varepsilon^{4}\delta_{2}^{2} + \mathcal{O}(\varepsilon^{5})
  \right].          
  \label{ede9}
\end{equation}

Using \eqref{ede7}, we have
\begin{align}
  \|\bb^1_{ij}\times\bb^2_{ij}\|^{2}
  &=
  \varepsilon^{4}\delta_{2}^{4}
  \biggl[
    1 + 2(\xi_{1,1}-\xi_{2,2})\varepsilon
  +
  \biggl(
    \eta_{,1}^{2}+\eta_{,2}^{2}+(\xi_{1,11}+\xi_{2,22})\delta_{2}
  \nonumber \\
  &
    +2(\xi_{1,1}\xi_{2,2}-\xi_{1,2}\xi_{2,1})
    +(\xi_{1,1}+\xi_{2,2})^{2}
  \biggr)\varepsilon^{2}
  + \mathcal{O}(\varepsilon^{3})  
  \biggr].   \label{ede10}
\end{align}
and
\begin{equation}
  \|\bb^1_{i-1j+1}\|^{2}
 =
  \varepsilon^{2}\delta_{2}^{2}
  \left[
    1 + 2\xi_{1,1}\varepsilon
   +
  \left(
    \xi_{1,1}^{2}+\xi_{2,1}^{2}
    +(2\xi_{1,12}-\xi_{1,11})\delta_{2}
    +\eta_{,1}^{2}
  \right)\varepsilon^{2}
  + \mathcal{O}(\varepsilon^{3})  
  \right].  \label{ede11}
\end{equation}

From \eqref{ede10} and \eqref{ede11} one checks that
\begin{align}
  \left[\|\bb^1_{ij}\times\bb^2_{ij}\|^{2}\|\bb^1_{i-1j+1}\|^{2}\right]^{-1}
  &=
  \left[ \varepsilon^{6}\delta_{2}^{6} (1+\mathcal{O}(\varepsilon)) \right]^{-1}
  =
  \varepsilon^{-6}\delta_{2}^{-6} (1+\mathcal{O}(\varepsilon)).
  \label{ede12}
\end{align}
Finally, combining \eqref{ede9} and \eqref{ede12}, we get the expansion
\begin{equation}
  \frac{{\left((\bb^1_{ij}\times\bb^2_{ij})\cdot\bb^1_{i-1j+1}\right)}^2}{{\|\bb^1_{ij}\times\bb^2_{ij}\|}^2{\|\bb^1_{i-1j+1}\|}^2}
  =
  (\eta_{,12}-\eta_{,11})^{2}\delta_{2}^{2}\varepsilon^{4}
  +
  \mathcal{O}(\varepsilon^{5}).
  \label{ede13}
\end{equation}
A similar computation for the second term in \eqref{e_dih} yields
\begin{equation}
  \frac{{\left((\bb^1_{ij}\times\bb^2_{ij-1})\cdot\bb^1_{i-1j-1}\right)}^2}
       {{\|\bb^1_{ij}\times\bb^2_{ij-1}\|}^2{\|\bb^1_{i-1j-1}\|}^2} 
  =
  (\eta_{,12}+\eta_{,11})^{2}\delta_{2}^{2}\varepsilon^{4}
  +
  \mathcal{O}(\varepsilon^{5}).
  \label{ede14}
\end{equation}
Likewise, for the first and fourth terms in \eqref{e_dih}, we have
   \begin{equation}
  \frac{{\left((\bb^2_{ij}\times\bb^1_{i-1j})\cdot\bb^2_{i-1j-1}\right)}^2}
       {{\|\bb^2_{ij}\times\bb^1_{i-1j}\|}^2{\|\bb^2_{i-1j-1}\|}^2}
  =
  (\eta_{,12}-\eta_{,22})^{2}\delta_{2}^{2}\varepsilon^{4}
  +
  \mathcal{O}(\varepsilon^{5})
  \label{ede15}
\end{equation}
 and 
\begin{equation}
  \frac{{\left((\bb^1_{i-1j}\times\bb^2_{ij-1})\cdot\bb^2_{i-1j}\right)}^2}
       {{\|\bb^1_{i-1j}\times\bb^2_{ij-1}\|}^2{\|\bb^2_{i-1j}\|}^2}
  = 
  (\eta_{,12}+\eta_{,22})^{2}\delta_{2}^{2}\varepsilon^{4}
  +
  \mathcal{O}(\varepsilon^{5}),
  \label{ede15pr}
  \end{equation}
respectively.

By combining \eqref{ede13}--\eqref{ede15pr}, we get
\begin{equation}
   \mathcal E_d[\boldsymbol\xi,\eta]:=
  \sum_{i,j=1}^{N_2} \frac{k_d}{\omega}
  \left[
    \eta_{,11}^2+2\eta_{,12}^2+\eta_{,22}^2
    \right]\delta_{2}^{2}\varepsilon^{5}
  +
  \mathcal{O}(\varepsilon^{6}).  
  \label{ede27}
\end{equation}

\bibliographystyle{ieeetr}
\bibliography{gamma-converg-references}

\end{document}